\documentclass[preprint,12pt]{elsarticle}

\usepackage{graphicx} 	  
\usepackage{amssymb} 	

\usepackage{color}
\usepackage{amsmath}
\usepackage{mathtools}
\usepackage{fullpage}
\usepackage{algorithmic}

\algsetup{linenosize=\small}
\usepackage[table,xcdraw]{xcolor}
\usepackage{multirow}
\usepackage[super]{nth}
\usepackage{graphicx,adjustbox}
\usepackage{caption}
\usepackage[labelformat=simple]{subcaption}
\usepackage{comment}
\usepackage{setspace}
\usepackage{textcomp}
\usepackage{xspace}
\usepackage{siunitx}
\usepackage{epsfig}
\usepackage{epstopdf}
\usepackage{soul}
\usepackage{url}
\usepackage{tablefootnote}
\usepackage[linesnumbered,ruled]{algorithm2e}
\usepackage{booktabs}
\usepackage{hyperref}
\usepackage[normalem]{ulem}
\usepackage{footnote}
\usepackage[misc,geometry]{ifsym} 
\usepackage{natbib}
\setcitestyle{authoryear,open={((},close={))}}
\makesavenoteenv{tabular}

\let\OldTexttrademark\texttrademark
\renewcommand{\texttrademark}{\OldTexttrademark\xspace}%

\usepackage[toc,page]{appendix}

\usepackage{epsfig}
\usepackage{amssymb}

\journal{Network and Computer Applications}

\begin{document}

\begin{frontmatter}

\title{Potential and Pitfalls of Multi-Armed Bandits for Decentralized Spatial Reuse in WLANs}

\author[label1]{Francesc Wilhelmi \corref{cor1}} 
\author[label1]{Sergio~Barrachina-Mu\~noz}
\author[label1]{Boris~Bellalta} \author[label2]{Cristina~Cano}
\author[label3]{Anders~Jonsson}
\author[label3]{Gergely~Neu}
\address[label1]{Wireless Networking Research Group (WN-UPF), 08002 Barcelona, Spain}
\address[label2]{Wireless Networks Research Group (WINE-UOC), 08860 Castelldefels (Barelona), Spain}
\address[label3]{Artificial Intelligence and Machine Learning Research Group (AIML-UPF), 08002 Barcelona, Spain}
\cortext[cor1]{Corresponding Author: Francesc Wilhelmi, Email address: \href{francisco.wilhelmi@upf.edu}{francisco.wilhelmi@upf.edu}}

\begin{abstract}
Spatial Reuse (SR) has recently gained attention to maximize the performance of IEEE 802.11 Wireless Local Area Networks (WLANs). Decentralized mechanisms are expected to be key in the development of SR solutions for next-generation WLANs, since many deployments are characterized by being uncoordinated by nature. However, the potential of decentralized mechanisms is limited by the significant lack of knowledge with respect to the overall wireless environment. To shed some light on this subject, we show the main considerations and possibilities of applying online learning to address the SR problem in uncoordinated WLANs. In particular, we provide a solution based on Multi-Armed Bandits (MABs) whereby independent WLANs dynamically adjust their frequency channel, transmit power and sensitivity threshold. To that purpose, we provide two different strategies, which refer to selfish and environment-aware learning. While the former stands for pure individual behavior, the second one considers the performance experienced by surrounding networks, thus taking into account the impact of individual actions on the environment. Through these two strategies we delve into practical issues of applying MABs in wireless networks, such as convergence guarantees or adversarial effects. Our simulation results illustrate the potential of the proposed solutions for enabling SR in future WLANs. We show that substantial improvements on network performance can be achieved regarding throughput and fairness.
\end{abstract}


\end{frontmatter}

\newpage

\section{Introduction}
\label{section:introduction}
Wireless communications are rapidly evolving to satisfy the increasingly tighter requirements coming from the explosive growth of wireless devices. To solve that, future Wireless Networks (WNs) are foreseen to cover small areas in high-density scenarios, which evidences the need for novel mechanisms to maximize spectral efficiency. In particular, Spatial Reuse (SR) has gained attention in recent years as a potential solution to improve the use of the spectrum. One of the most prominent examples can be found in the IEEE 802.11ax-2019 (11ax) amendment (\citealp{bellalta2016ax}), which defines High-Efficiency (HE) Wireless Local Area Networks (WLANs). The 11ax amendment aims to maximize spectral efficiency through the SR operation and other spectrum-efficient techniques like Orthogonal Frequency-Division Multiple Access (OFDMA), and Uplink/Downlink Multi-User Multiple-Input-Multiple-Output (MU-MIMO).

In this paper, we focus on IEEE 802.11 WLANs, which mostly represent uncoordinated deployments (e.g., residential buildings). These networks are limited in performance because of the scalability issues arising from the current decentralized channel access mechanisms, i.e., Carrier Sense Multiple Access with Collision Avoidance (CSMA/CA) (\citealp{ergin2007understanding}). To enable SR in WLANs, we consider the use of Transmit Power Control (TPC) and Carrier Sense Threshold (CST) adjustment. These mechanisms are of particular concern to the 11ax amendment. They facilitate and regulate the SR operation by providing a set of procedures and constraints for dynamically setting the transmit power and the sensitivity. Roughly, the idea of TPC lies in adjusting the transmit power for reducing the interference and/or saving energy. Similarly, CST adjustment seeks to increase the number of parallel transmissions by modifying the sensitivity on a per-device basis. In the context of IEEE 802.11ax, the Dynamic Sensitivity Control (DSC) has been proposed as a potential solution for enabling SR through sensitivity adjustment (\citealp{smith2015dynamic}).

However, addressing the SR problem in WLANs through TPC and CST adjustment is not trivial for decentralized deployments. This is mostly motivated by \emph{i)} the spatial interactions between nodes, and \emph{ii)} the adversarial setting unleashed in decentralized wireless networks. For the former, tuning either the transmit power or the sensitivity entails dealing with the spatial dimension. Unlike for frequency and temporal approaches, spatial interference cannot be treated as a binary model. In the latter case, one can observe when devices transmit or not, thus obtaining full or null performance respectively. As shown later in Section \ref{section:interactions_wlans}, more complex interactions occur between WLANs tuning their transmit power and CST. As a result, modeling inter-WLANs interactions in next-generation deployments turns out to be extremely complex. Moreover, regarding the adversarial setting unleashed by decentralized deployments, strong competition between independent networks may occur. 

In order to address the SR problem in high-density decentralized networks, we focus on the Multi-Armed Bandit (MAB) framework. The MAB approach frames the learning-by-interaction problem, and allows to properly approach the exploration-exploitation trade-off in face of uncertainty. In MABs, a learner (or agent) obtains information from the environment, which reacts according to the actions performed - in the SR problem, an action may refer to a certain configuration of transmit power and CST. By interacting with the environment, a given learner aims to maximize a numerical cumulative reward over time. Unlike classical Reinforcement Learning (RL), the MAB setting does not consider states in general.\footnote{There is a class of bandits problems that consider states (stateful bandits), which is not considered due to the characteristics of the problem addressed in this work. Essentially, in this class of bandits, states are usually represented by taking strong assumptions that hinder the accuracy of the analysis.} A state may refer to a concrete temporal situation in which the learner is involved. Therefore, it allows the latter to construct a policy that determines the behavior to be followed in future situations. Accordingly, learning through states adds an extra layer of complexity and requires that a given agent learns additional contextual information.

The application of MABs into wireless communications problems has recently become very popular (\citealp{chen2010distributed, maghsudi2015channel, maghsudi2015joint}). To model the SR problem through MABs, we consider that WLANs are empowered with an agent that attempts to learn the best-performing action (i.e., a combination of the frequency channel, the transmit power and the sensitivity level). Its learning operation is based on the performance achieved in an unknown environment. In this way, MABs operate on the top of CSMA/CA, which operation is influenced by the spatial interactions generated by the taken actions. As a result, we expect WLANs to autonomously find their best configuration in an adversarial setting, given a performance maximization strategy.

The main goal of this paper, then, is to determine the feasibility of applying decentralized learning to improve spectral efficiency in next-generation wireless deployments. In particular, we apply online learning mechanisms to enable SR in dense and uncoordinated WLANs, and show the main derived implications and considerations. We highlight the impact on the aggregate performance and fairness experienced by WLANs, as well as on the guarantees for converging to the optimal solution. The implications of applying online learning to WLANs are studied through the utilization of selfish and environment-aware learning-based strategies. While a selfish strategy is based on the individual performance of a given learner, environment-aware considers a set of neighboring WLANs that share a reward. The SR problem presented in this work is non-convex, therefore prevents to provide any kind of convergence bound. However, our results suggest that the performance of WLANs can be maximized by using learning strategies that are based on probabilistic models of the reward. To summarize, the main contributions of this work are described next:
\begin{itemize}
	\item We showcase the major inter-WLAN dependencies when modifying both the transmit power and the CST, and how they affect the network performance.
	\item In order to capture the CSMA/CA operation of IEEE 802.11 WLANs, we use Continuous Time Markov Networks (CTMNs) in spatially-distributed scenarios (\citealp{bellalta2017throughput}). We show that CTMNs models are able to capture the existing dependencies between overlapping WLANs. 	
	\item We model the SR problem in WLANs through MABs, where agents implementing Thompson sampling decide the configuration of a given network in terms of frequency channel, transmit power and CST. To the best of our knowledge, this is the first work applying MABs on a CSMA-based network.
	\item We provide insights on the main considerations of using learning in decentralized and adversarial wireless networks. In particular, we showcase the implications of applying selfish and environment-aware learning in dense WLANs, thus emphasizing on the main potentials and pitfalls.
	\item We evaluate the performance of using self-configuring agents in dense WLANs, both in specific and random scenarios. The two learning approaches presented in this paper are shown to significantly improve the performance achieved by WLANs in terms of throughput and fairness, with respect to a default - and static - configuration.
\end{itemize}		

The remaining of this document is structured as follows: Section \ref{section:previous_work} refers to the previous related work. In Section \ref{section:interactions_wlans} we first provide details on the throughput model considered in this paper to fit the SR problem. Then, we characterize inter-WLAN interactions when tuning both the transmit power and the sensitivity threshold. A set of illustrative scenarios is used for that purpose. Section \ref{section:mabs} formulates the SR problem through MABs and shows the main implications to be considered when applying learning to decentralized wireless networks. Then, Section \ref{section:performance_evaluation} shows the main results of this paper with regard to selfish and environment-aware learning in WLANs. Finally, some remarks are given in Section \ref{section:conclusions}.

\section{Related Work}
\label{section:previous_work} 		
Machine Learning (ML), and more precisely RL, has received increasing interest from the wireless communications research community over the last years. One of the main reasons resides in the increased complexity of problems related to next-generation wireless systems. Such kind of environments are characterized by being particularly dense, so that the best configuration strategy may be difficult to foresee. Since the current preprogrammed approaches are likely to be suboptimal, RL is expected to improve the action selection from experience. In particular, RL-based methods are expected to provide close-to-optimal solutions to complex problems within an acceptable timescale, which is an indispensable requirement in wireless networks.

To the best of our knowledge, one of the first works to apply RL into a SR-related problem in wireless networks is (\citealp{nie1999q}), in which the authors show a centralized Q-learning mechanism to dynamically select the channel in mobile networks. Other RL-based approaches for channel access can be found in (\citealp{li2009multi, bennis2010q, bennis2011distributed, sallent2015learning, rupasinghe2015reinforcement}), covering cognitive radio, self-organizing cellular networks and coexistence problems. Despite Q-learning (or other Markovian-based methods) has been shown to properly fit to channel allocation problems, few applications have been provided to the SR problem. Note, as well, that dealing with the frequency domain allows to naturally define states,\footnote{An state describes a particular situation of a given agent at a specific time.} which can be done according to the availability of channels (typically modeled through Bernoulli distributions). Therefore, an agent may observe the environment and define an accurate model where the state is defined by the set of channels that are available/occupied. Note that the contextual information provided to the learner is important for learning efficiently, since the agent is able to react to different situations. With a proper definition of states, a higher degree of control is conferred to the agent. Therefore, provided that the model of the states is accurate enough, the learning procedure carried out by a given learner can result into better performance than that of a stateless setting. 

However, modeling states for the decentralized SR problem entails added complexity, thus hindering the learning procedure followed by a given agent. In the particular case of IEEE 802.11 WLANs, spatial interactions among nodes lead to complex scenarios. The performance achieved by a given WLAN depends on the additive interference coming from an unknown environment. Therefore, learning accurate enough states for the SR problem turns out to be challenging. Note that, if states do not reflect the actual situation of a given agent at a given moment, the learnings that can be generated become strongly limited, and can even be meaningless. To cope with the difficulties on modeling states for the decentralized SR problem, we focus on multi-player MABs (MP-MABs), which frame resource allocation problems where several agents compete against each other. MP-MABs have been recently broadly applied for opportunistic spectrum access in cognitive radio (\citealp{liu2010distributed, anandkumar2011distributed, rosenski2016multi, maghsudi2015joint, maghsudi2015channel}). 

Firstly, in (\citealp{liu2010distributed}), the authors provide a decentralized policy with logarithmic regret order, which is based on a time-division fair sharing of the best arms. However, such a policy requires coordination among agents and to know the exact number of adversarial nodes, which in addition must be constant and known in advance. Both requirements entail dedicated communication between nodes, which turns out to be unfeasible for decentralized problems such as the one presented in this paper. Another important contribution regarding multi-player learning for the opportunistic spectrum access problem is provided in (\citealp{anandkumar2011distributed}), where the authors provide a distributed learning algorithm that showcases order-optimal regret. However, the total number of secondary users is known by the system, which may not be feasible in real scenarios where no communication between nodes exists. In contrast, in this work we consider selfish and environment-aware learning approaches, none of which require explicit communication between independent learners. Furthermore, a less strict method is also provided in (\citealp{anandkumar2011distributed}), in which the number of secondary users (which is fixed) is estimated, so that nearly order-optimal regret is achieved. In both algorithms, it is assumed that all the users use the same policy. Regarding the work in (\citealp{rosenski2016multi}), sublinear regret is achieved if all players implement the proposed algorithm. Some interesting thoughts are provided regarding varying environments, which, to the best of our knowledge, have been barely considered in the previous literature. For instance, the authors emphasize that, in the dynamic setting, the frequency at which players enter and leave the scenario must be limited in order to provide a sublinear regret. Unlike the SR presented in this paper, the work in (\citealp{rosenski2016multi}) assumes that there exists an optimal solution whereby no collisions occur for any player. The concept of collision is inspired in the ALOHA channel access mechanism, and occurs if two or more players choose the same arm (or channel). 
 
When it comes to the SR problem by means of TPC and/or sensitivity adjustment, we find MP-MABs application for joint channel selection and power control in (\citealp{maghsudi2015joint, maghsudi2015channel}). The work in (\citealp{maghsudi2015joint}) proposes a strategy based on the Signal-to-Interference-plus-Noise Ratio (SINR) to determine the channel and the transmit power to be used in cognitive radio networks. The authors prove that the MP-MAB game converges to a correlated equilibrium, which in addition maximizes the aggregate utility, if the two following assumptions hold: \emph{i)} the problem is relaxed, so that the reward granted to a given agent depends only on the Signal-to-Noise Ratio (SNR), regardless of the overlapping interference, \emph{ii)} user-specific penalties are provided to each agent. In contrast with the work in (\citealp{maghsudi2015joint}), here we aim to understand the potentials and limitations of applying MABs in a CSMA/CA-based setting, in which the previous assumptions do not hold. First of all, instead of relaxing the problem to only consider the SNR, we model the interactions at the MAC level. Secondly, defining user-specific penalties would require the use of either a centralized system or message exchanging.

Finally, in (\citealp{maghsudi2015channel}), the authors introduce the concept of calibrated forecaster, i.e., a predictor of the actions of the adversaries that improves with collected knowledge. By using such a predictor, if every learner is able to predict and respond to the others' actions, then the game converges to a correlated equilibrium. In other words, if a node can predict which channel will the adversary pick (and vice-versa), then it can select the other channel and experience the maximum performance. In contrast, in the SR problem we tackle here, developing an accurate forecaster in a decentralized way may be an extremely complex task. Refer to the non-linear relationships that occur in the spatial domain, which are hard to model, and thus to predict. Moreover, density and messy deployments may remove the existence of an equilibrium, hence invalidating the assumptions.

In summary, a lot of effort has been recently made to enable the evolution of wireless networks towards self-adjusting systems. In particular, the application of RL has been extensively studied for channel access problems. However, these type of methods do not properly suit to spatially-distributed problems such as the SR one. As a result, other stateless techniques, such as MABs, have been targeted, and have shown to effectively improve wireless networks performance, even in adversarial environments. Nevertheless, these mechanisms require that strong assumptions about the system model hold. Moreover, spatial interactions between WLANs have not yet been considered.

\section{Interactions between WLANs when Spatial Reuse is Enabled}
\label{section:interactions_wlans}    
In this Section, for completeness, we first briefly introduce the CSMA/CA operation carried out by Wi-Fi networks for accessing the channel, as well as the CSMA/CA throughput model considered along this paper. Therewith, we aim to identify the main inter-WLAN interactions when modifying both the transmit power and the CST. Understanding these interactions is key to motivate the usage of MABs to the decentralized SR problem. As shown in Section \ref{section:previous_work}, some of the previous work addressed similar problems and provided mechanisms that were proven to converge to an equilibrium. However, the novelty of this paper lies in the analysis of learning techniques in CSMA/CA-based networks. Unlike previous work, where the reward (i.e., the throughput) is mostly given by a linear function that only depends on the signal strength and the interference, here we deal with more complex interaction between networks. In Wi-Fi, due to the decentralized nature of CSMA/CA, an optimal solution in terms of SR is harder to derive than in cellular-based networks. In addition, there is a trade-off between performance maximization and fairness, which is not trivial to compute in a decentralized setting.
	
\subsection{CSMA/CA}
\label{section:csma}	
\begin{figure}[h!]
	\centering		
	\begin{subfigure}[b]{0.25\textwidth}
		\includegraphics[width=\textwidth]{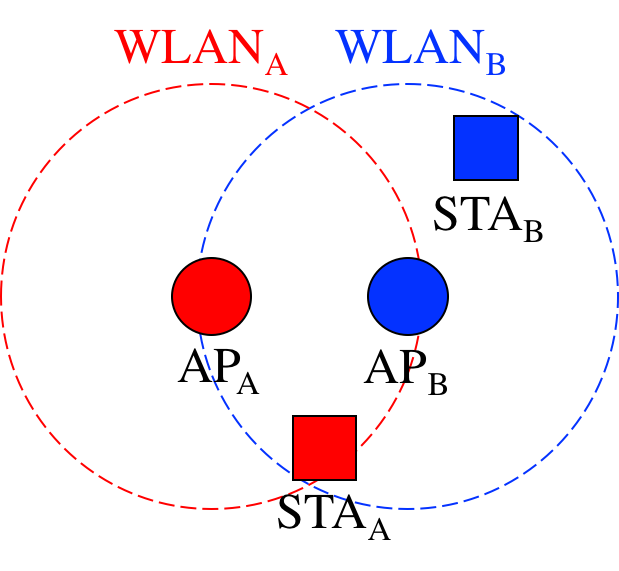}
		\caption{Scenario}\label{fig:csma_a}
	\end{subfigure}
	\begin{subfigure}[b]{0.55\textwidth}
		\includegraphics[width=\textwidth]{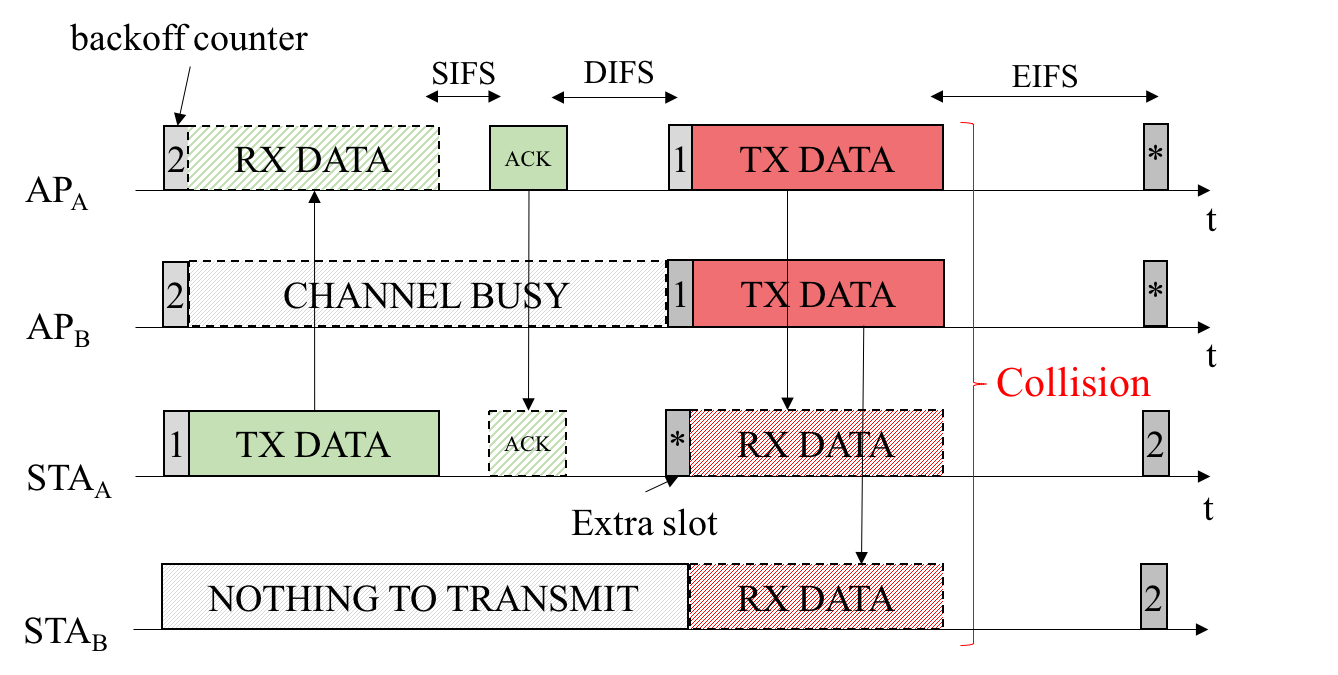}
		\caption{CSMA/CA operation}\label{fig:csma_b}
	\end{subfigure}
	\caption{CSMA/CA operation in $\text{WLAN}_{\text{A}}$ and $\text{WLAN}_{\text{B}}$. $\text{STA}_{\text{A}}$ starts a transmission to $\text{AP}_{\text{A}}$, since its backoff counter reaches zero first. After that, a collision occurs due to simultaneous transmissions held by $\text{AP}_{\text{A}}$ and $\text{AP}_{\text{B}}$.}
	\label{fig:csma}
\end{figure}

Channel access is performed in IEEE 802.11 WLANs by means of the Distributed Coordination Function (DCF), which is based on CSMA/CA. In DCF, before being able to transmit a packet, a transmitter must listen to the channel for a period of time called Distributed Inter Frame Space (DIFS). The channel is sensed to be free according to the Clear Channel Assessment (CCA) mechanism, i.e., if the power perceived is lower than a given threshold.\footnote{Throughout this paper, we refer to CCA and CST indistinctly.} The power received at a given node is the sum of all the interference generated by the other devices under the environment-constrained propagation effects. Furthermore, the access to the medium is randomized in order to reduce the number of potential collisions between other contenders. Specifically, each transmitter selects a random backoff value to start a countdown that is active as long as the channel remains free. In case the channel is sensed busy, the countdown is paused. It is resumed as soon as the ongoing transmission finishes and the channel is sensed free again. An example of the CSMA/CA operation is shown in Figure \ref{fig:csma}, where we show two overlapping WLANs, namely $\text{WLAN}_{\text{A}}$ and $\text{WLAN}_{\text{B}}$, respectively. Station A ($\text{STA}_{\text{A}}$) is the first to gain access to the channel, so it starts a transmission to the Access Point A ($\text{AP}_\text{A}$). Meanwhile, $\text{AP}_{\text{B}}$ senses the channel busy and freezes its backoff. After this first transmission, both $\text{AP}_{\text{A}}$ and $\text{AP}_{\text{B}}$ access to the channel simultaneously because they both randomly chose the same backoff counter. As a result, a collision is produced.

\subsection{CSMA/CA Throughput Model}
\label{section:throughput_model}		
For the rest of this paper we consider that WLANs are independent entities composed by an AP and a STA, in which saturation downlink traffic (i.e., from the AP to the STA) is assumed. Such an assumption is  reasonable as long as we target home deployments, where STAs are often very close to the AP. Moreover, the main goal of this paper is to capture inter-WLAN interactions.

Regarding the throughput model, we rely on the CTMN-based analytical tool for spatially distributed WLANs presented in (\citealp{barrachina2018performance}), referred to as SFCTMN. This tool captures the interrelations given in scenarios where nodes operating in the same channel are not required to be within the carrier sense range of each other. Essentially, given a scenario (i.e., nodes' location, channels, transmission powers, CCA levels, path loss model, etc.), states and transitions are generated in accordance with the CSMA/CA mechanism. That is, WLANs are only allowed to decrement their backoff and start transmissions when the CCA condition is accomplished. In (\citealp{barrachina2018overlap}), authors use SFCTMN to assess the performance of high density WLANs under different traffic loads.

\begin{figure}[h!]
	\centering		
	\begin{subfigure}[b]{0.35\textwidth}
		\includegraphics[width=\textwidth]{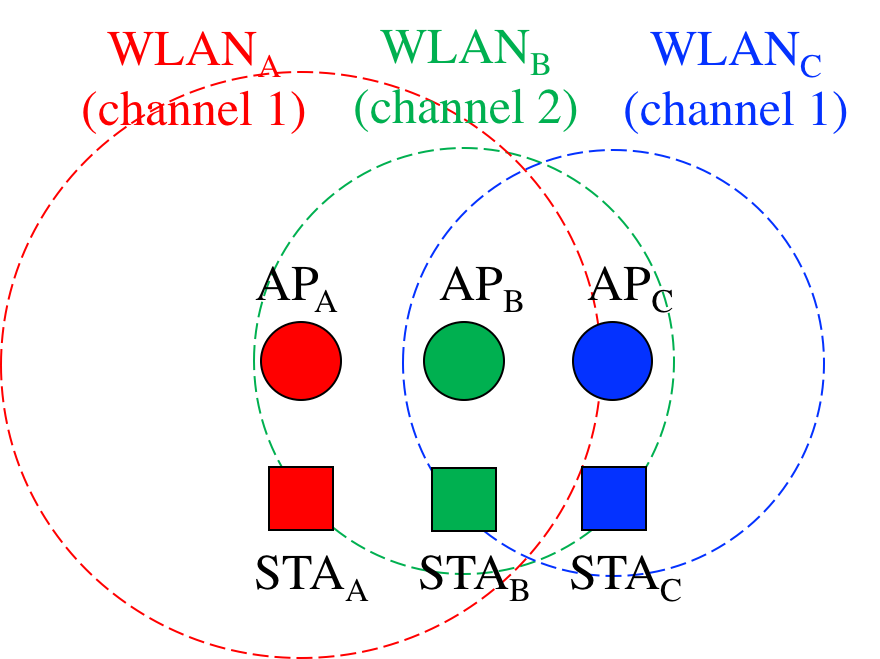}
		\caption{Scenario with overlapping WLANs}\label{fig:three_wlans_ctmn_a}
	\end{subfigure}
	\begin{subfigure}[b]{0.5\textwidth}
		\includegraphics[width=\textwidth]{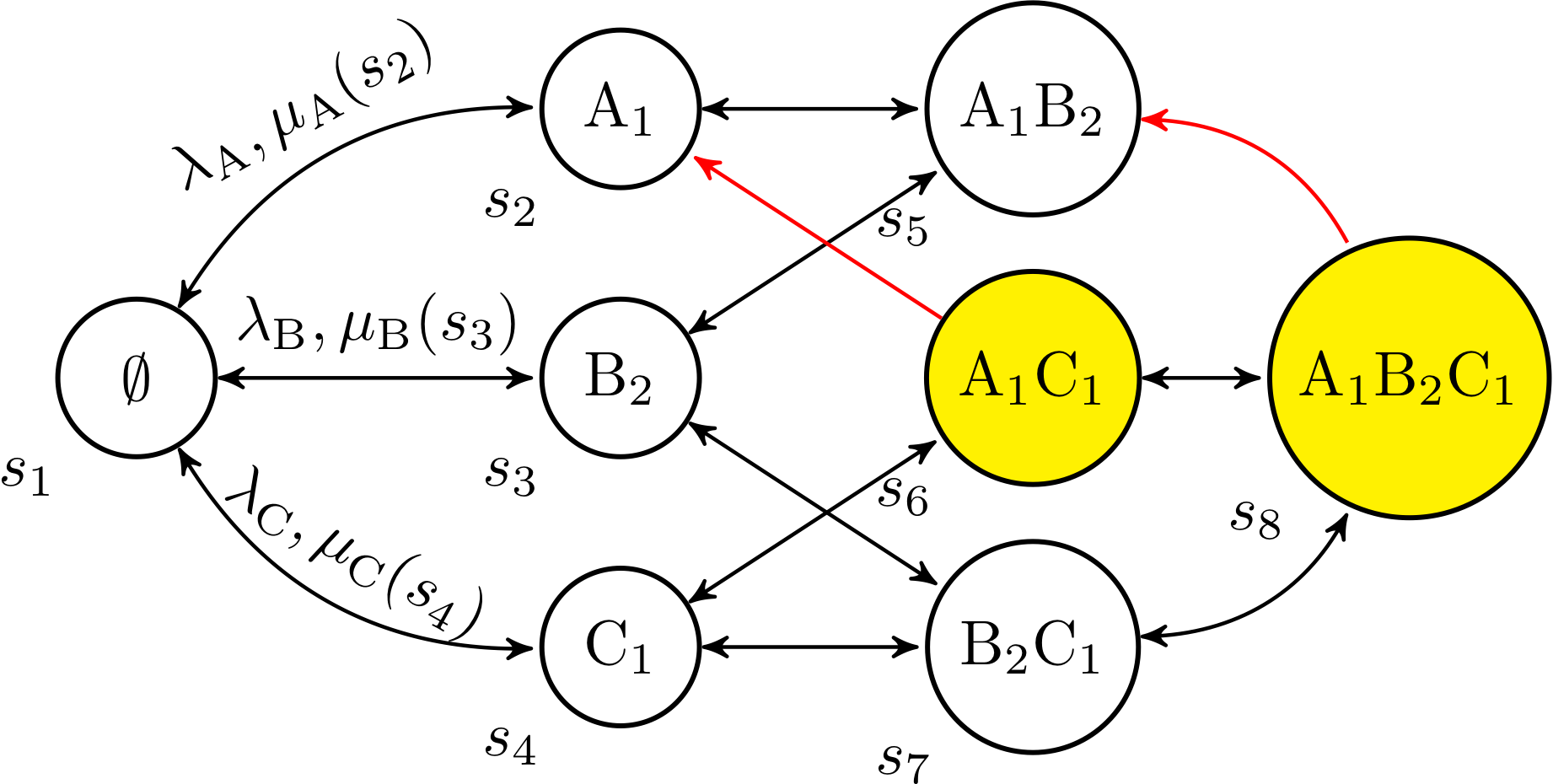}
		\caption{Markov chain}\label{fig:three_wlans_ctmn_b}
	\end{subfigure}
	\caption{Toy scenario. WLANs A and C operate in channel 1 while B operates in channel 2. Note that C is in the carrier-sense range of A. Only the transition rate pairs ($\lambda, \mu$) between states $s_1$, $s_2$, $s_3$ and $s_4$ are displayed for the sake of visualization. The states where C may lose packets because of the interference from A are displayed in yellow. Unidirectional backward transitions are show in red.}    
	\label{fig:three_wlans_ctmn}
\end{figure}

A state in the CTMN is defined by the set of WLANs active and the channel in which they are transmitting.\footnote{Note that in this work we assume only 20 MHz single-channel transmissions.} Accordingly, transitions between states occur if WLANs become active/disabled. For example in state $\text{A}_1 \text{B}_2 \text{C}_1$ there are three active WLANs: A, B and C transmit in channels 1, 2 and 1, respectively. Since states and transitions are generated according to the regular CSMA/CA mechanism, a CTMN may have both bidirectional and unidirectional transitions between states. It is the case of the toy scenario shown in Figure \ref{fig:three_wlans_ctmn}, where A uses a higher transmission power than B and C. While A is able to access channel 1 when C is transmitting, C is not able to do so when A is transmitting because of the high interference sensed in channel 1. Accordingly, only backward transitions are permitted from state $s_6 = \text{A}_1 \text{C}_1$ to $s_2 = \text{A}_1$, and from state $s_8 = \text{A}_1 \text{B}_2 \text{C}_1$ to $s_6 = \text{A}_1 \text{B}_2$. Essentially, given the channel and power configurations of this particular scenario, while A operates like in isolation, C's operation is subject to A's behavior. Note that B also operates like in isolation since it uses a different channel.

Transitions between any two states $s$ and $s'$ in the CTMN have a corresponding transition rate $R_{s,s'}$. For \textit{forward} transitions (i.e., when a WLAN starts a new transmission), the average packet transmission attempt rate is $\lambda = 1/\text{E}[B]$, being $\text{E}[B]$ the expected backoff duration. For \textit{backward} transitions (i.e., when a WLAN finishes its transmission), the departure rate ($\mu$) depends on the duration of a transmission. The latter is given by both the data rate (subject to the selected Modulation and Coding Scheme (MCS) and transmission channel width) and the average data packet length ($\text{E}[L]$). Thus, we simply say that the data rate of a WLAN $w$ depends on the state $s$ of the system, $\mu_{w}(s)$, in other words, on the set of overlapping WLANs that transmit simultaneously. The information contained in a given state, therefore, refers to the inter-WLAN interactions in that situation.

In order to estimate the average throughput experienced by each WLAN in a given scenario, we must first estimate the fraction of time the system spends in each state ($\vec{\pi}$). We define $\pi_s$ as the probability of finding the system at state $s$. In continuous-time Markov processes with stationary distribution, $\vec{\pi}$ is given by solving the system of equations $\text{Q} \vec{\pi} = 0$, where the matrix item $\text{Q}$ is the infinitesimal generator of the CTMN.
Given $\vec{\pi}$, the average downlink throughput of WLAN $w$ in a given state $s$ can be defined as
\begin{equation*}
\Gamma_{w}(s) := \begin{cases} 
\text{E}[L]  \mu_w(s) \pi_s  \text{,}  & \text{SINR}_{w}(s) > \text{CE} \\
0 \text{,} & \text{otherwise}
\end{cases}
\end{equation*}
where $\text{SINR}_{w}(s)$ is the SINR perceived by the receiving STA in WLAN $w$ in state $s$, and CE is the capture effect threshold.	Therefore, the resulting average downlink throughput that a given WLAN $w$ experiences can be computed as $\Gamma_w = \sum_{s \in \mathcal{S}}^{}\Gamma_{w}(s) $.

\subsection{Analysis}
\label{section:spatial_reuse_enhancement}  
\begin{figure}[t!]
	\centering		
	\begin{minipage}[b]{0.35\textwidth}		
		\begin{subfigure}[b]{0.8\linewidth}
			\includegraphics[width=\textwidth]{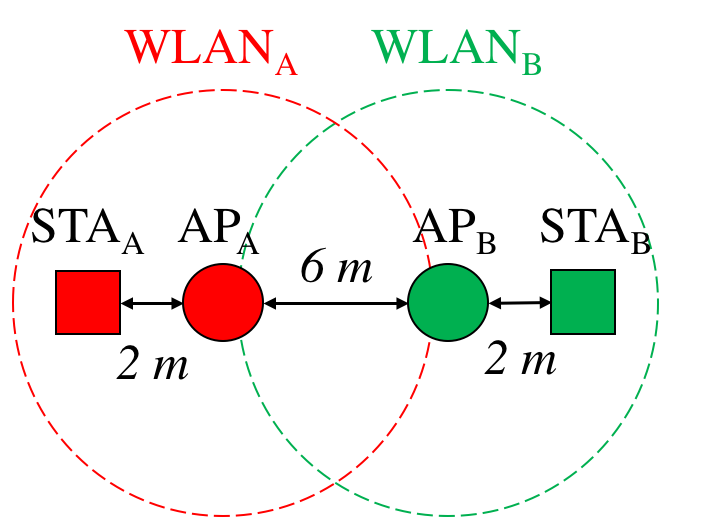}
			\caption{Scenario 1}\label{fig:s1_interactions}
		\end{subfigure}\\[\baselineskip]
		\begin{subfigure}[b]{\linewidth}
			\includegraphics[width=\textwidth]{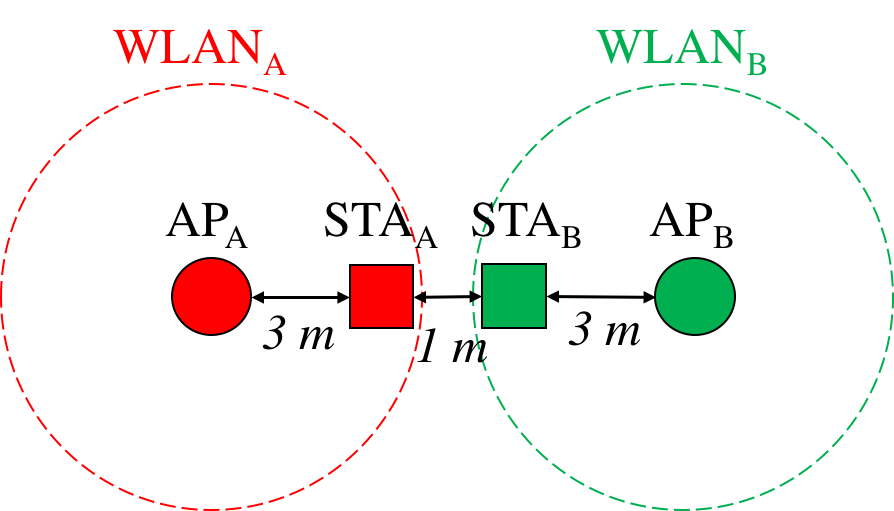}
			\caption{Scenario 2}\label{fig:s2_interactions}
		\end{subfigure}
	\end{minipage}
	\begin{minipage}[b]{0.3\textwidth}		
		\begin{subfigure}[b]{\linewidth}
			\centering	
			\includegraphics[width=\textwidth]{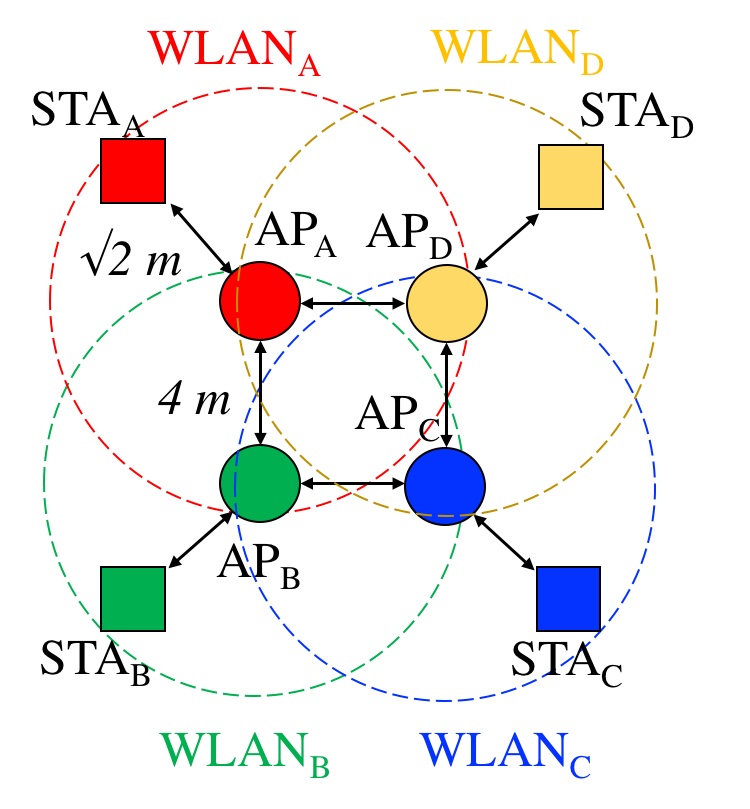}
			\caption{Scenario 3}\label{fig:s3_interactions}
		\end{subfigure}\\[\baselineskip]
	\end{minipage}
	\begin{subfigure}[b]{0.6\linewidth}
		\centering	
		\includegraphics[width=\textwidth]{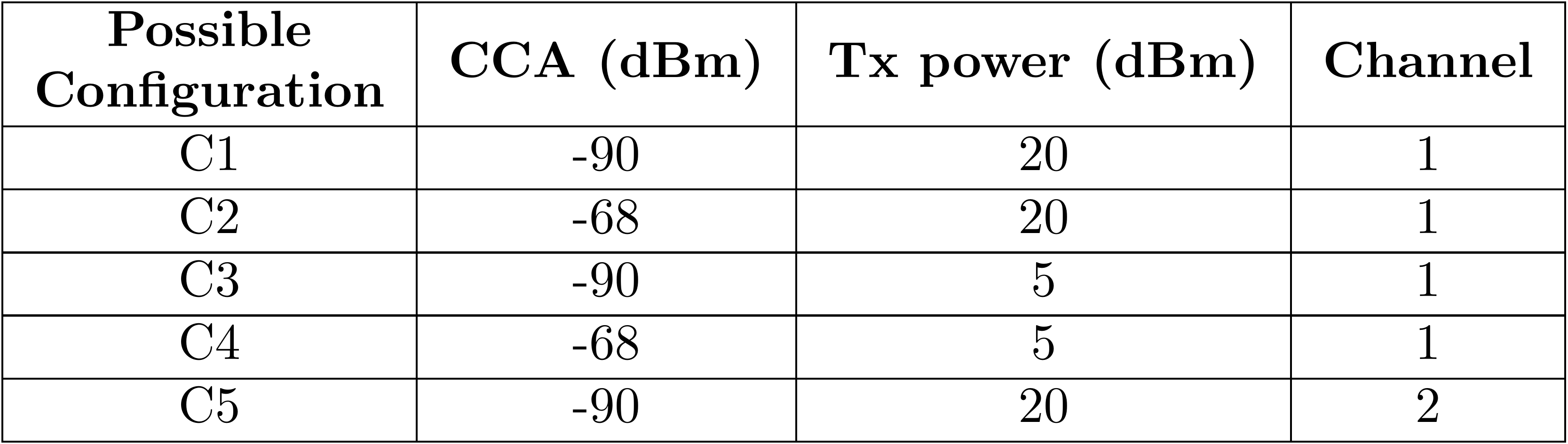}
		\caption{Table with possible configurations to be chosen by WLANs}\label{tbl:configurations}
	\end{subfigure}		
	\caption{Scenarios for characterizing inter-WLAN interactions.}
	\label{fig:scenarios_validations}
\end{figure}

To underline the potential of adjusting both the transmit power and the CST to enable SR in overlapping WLANs, we next introduce the main performance issues and anomalies that characterize IEEE 802.11 networks. Before, and in order to further analyze these issues, we introduce the set of scenarios shown in Figure \ref{fig:scenarios_validations}. This set of scenarios is evaluated under different static configurations (shown in Table \ref{tbl:configurations}), each one referring to a specific combination of channel, transmit power and CCA. Such combinations (from \emph{C1} to \emph{C5}) refer to specific configurations of the set of allowed values that a given WLAN can choose, which are detailed in \ref{section:simulated_wireless_environment}, together with simulation parameters. Results shown in Table \ref{tbl:results_sr_improvements} were obtained by applying the analytical model presented in Section \ref{section:throughput_model}\footnote{All of the source code used in this work is open (\citealp{fwilhelmi2018code}) under the GNU General Public License v3.0, encouraging sharing of knowledge between potential contributors.} and the \texttt{11axHDWLANSim} simulator.\footnote{The source code of \texttt{11axHDWLANSim} is open under the GNU General Public License v3.0 and can be found at \url{https://github.com/wn-upf/Komondor}}

\begin{table}[t!]
	\centering
	\begin{tabular}{|c|c|c|c|}
		\hline
		\textbf{Scenario} & \textbf{Conf.} & \textbf{$\overline{\Gamma}$ (SFCTMN)} & \textbf{$\overline{\Gamma}$ (\texttt{11axHDWLANSim})} \\ \hline
		\multirow{3}{*}{1} & C1 & 56.90 Mbps &  56.94 Mbps\\ \cline{2-4} 
		& C2 & 113.23 Mbps & 113.23 Mbps\\ \cline{2-4} 
		& C3 & 62.43 Mbps & 62.43 Mbps\\ \hline
		\multirow{2}{*}{2} & C2 & 0.73 Mbps & 0.00 Mbps\\ \cline{2-4} 
		& C4 & 62.43 Mbps & 62.43 Mbps \\ \hline
		\multirow{2}{*}{3} & C1 & 56.62 Mbps & 56.62 Mbps\\ \cline{2-4} 
		& C1 \& C5 & 113.23 Mbps & 113.23 Mbps \\ \hline
	\end{tabular}
	\caption{Performance in each scenario achieved through different configurations. Each cell contains the performance computed by using CTMNs and Komondor, respectively. Komondor results are extracted from 1,000 s simulations.}
	\label{tbl:results_sr_improvements}
\end{table}

\begin{itemize}		
	\item \textbf{Exposed-terminal problem:} two or more WLANs are not able to transmit simultaneously due to the inter-WLAN interference, which is higher than the CCA threshold at the transmitter. However, the receiver would be able to properly decode the data of interest, even in presence of other transmissions. In \emph{Scenario 1} (Figure \ref{fig:s1_interactions}), the exposed-terminal problem occurs if all the WLANs use configuration \emph{C1}. Such a situation is solved if WLANs apply configuration \emph{C2}, which consists in increasing the CCA in a way that both $\text{AP}_\text{A}$ and $\text{AP}_\text{B}$ can transmit simultaneously while using the same transmit power. In this case, both WLANs receive the same interference, but, by using a higher CCA, it is dismissed and does not force contention. Similarly, if WLANs reduce both the transmit power and the sensitivity (configuration \emph{C3}), the number of parallel transmissions can be increased. However, a lower performance compared to \emph{C2} is achieved due to the use of a lower MCS.		
	\item \textbf{Hidden-terminal problem:} occurs when two nodes that are not visible each other transmit simultaneously (not necessarily to the same destination), thus producing collisions. In particular, packet losses occur when the sensed interference at a given receiver results in a SINR lower than its capture effect. The hidden-terminal problem is framed in \emph{Scenario 2} (Figure \ref{fig:s2_interactions}) when both WLANs use configuration \emph{C1}. As a result, $\text{AP}_\text{A}$ and $\text{AP}_\text{B}$ can transmit simultaneously due to the CCA condition. However, if they do so, the SINR experienced at both $\text{STA}_\text{A}$ and $\text{STA}_\text{B}$ falls below their capture effect, thus leading to a wrong packet decoding. Such a situation is improved when $\text{AP}_\text{A}$ and $\text{AP}_\text{B}$ apply configuration \emph{C4}, which allows reducing the sensitivity area (higher CCA) and the generated interference (lower transmit power).		
	\item \textbf{Contending nodes:} similarly to the exposed-terminal problem, the channel is underutilized if one or more WLANs must defer their transmissions when another one is transmitting. In this case, increasing the CST and/or decreasing the transmit power in an appropriate manner may help at reducing the number of contending nodes. As a result, the number of parallel transmissions is maximized. This phenomena, in addition to be closely related to the exposed-terminal problem shown in \emph{Scenario 1}, can be further observed in \emph{Scenario 3} (Figure \ref{fig:s3_interactions}). In this case, in addition of using TPC and/or CST adjustment, we maximize SR by providing a proper channel allocation, which is done by combining configurations \emph{C1} and \emph{C5}. Hence, configurations are assigned so that frequency reuse is maximized. Moreover, there are situations that may require the opposite. That is to say, to force contention between nodes in order to prevent collisions. Such a situation occurs in \emph{Scenario 2} when configuration \emph{C2} is employed,\footnote{There is a significant difference in the throughput when applying \emph{C2} to \emph{Scenario 2} between the CTMNs model and \texttt{11axHDWLANSim} simulator. The fact is that CTMNs consider the time spent in each state. In this case, the dominant state is the one in which both WLANs transmit and experience collisions. However, the time spent in states where individual transmissions are held is considered, even if it is very small. In practice, transmissions affected by overlapping interference would result into null throughput, which is shown via the \texttt{11axHDWLANSim} simulator.} thus leading to zero throughput in both WLANs due to the collisions by hidden node. However, if a contending situation is forced by either increasing the transmit power or decreasing the CST (which occurs when setting configuration \emph{C4}), then the network performance is increased. 		
	\item \textbf{Flow starvation:} a given WLAN may be deprived of accessing the channel in case of noticing an excessive interference from other WLANs that do not sense each another. Such a phenomena can be solved by tuning both the transmit power and the CST. However, due to the nature of the problem, it may require some level of collaboration, since interfering nodes are completely unaware on the damage caused to the most vulnerable WLANs in terms of sensed interference. Flow starvation is studied in detail in Section \ref{section:mabs}.
	\item \textbf{Asymmetries:} finally, it is important to remark the consequences of existing asymmetries in a wireless network, which are mostly generated by the different situations of coexisting WLANs. The performance of a given WLAN is basically limited by its geographical location and possible configurations. Accordingly, there can be WLANs more privileged than others, so that the interference they sense is generally lower, thus experiencing a higher performance. Therefore, due to the spatial interactions generated by certain transmit power and CST levels, asymmetries may lead to a monopolization of the channel by dominant WLANs (i.e., enjoying better conditions than others). The effect of asymmetries is studied in detail in Section \ref{section:performance_evaluation}.	
\end{itemize}	 

As shown in the previous simulations, modifying either the transmit power or the CST in a WLAN may have severe implications on different communication aspects due to the utilization of CSMA/CA. While TPC allows to adjust the generated interference, CST adjustment aims to  modify the sensitivity area. It is worth to mention that SR can be enhanced if short-range communications are held, which can be achieved if using the minimum necessary transmit power and the maximum possible CST. Conversely, longer-range communications can be achieved when using a high transmit power and a low CST. Increasing the area of operation is useful to minimize performance issues such as flow starvation and collisions by hidden nodes. Table \ref{tbl:cca_tpc_effects} summarizes the intuitive effects of TPC and CST adaptation in WLANs.			
\begin{table}[h!]
	\centering
	\begin{tabular}{|c|c|c|c|}
		\hline
		\multirow{2}{*}{\begin{tabular}[c]{@{}c@{}}\\ \textbf{Action}\end{tabular}} & \multicolumn{3}{|c|}{\textbf{Effect}} \\ \cline{2-4} 
		& \begin{tabular}[c]{@{}c@{}}Exposed nodes\end{tabular} & \begin{tabular}[c]{@{}c@{}}Hidden nodes\end{tabular} & Data Rate \\ \hline
		$\uparrow$ Power & $\uparrow$ & $\downarrow$ & $\uparrow$ \\ \hline
		$\downarrow$ Power & $\downarrow$ & $\uparrow$ & $\downarrow$ \\ \hline
		$\uparrow$ CST & $\downarrow$ & $\uparrow$ & $\downarrow$  \\ \hline
		$\downarrow$ CST & $\uparrow$ & $\downarrow$ & $\uparrow$ \\ \hline
	\end{tabular}
	\caption{Effects of TPC and CST adjustment.}
	\label{tbl:cca_tpc_effects}
\end{table}

\section{Multi-Armed Bandits for Decentralized Spatial Reuse}
\label{section:mabs}
Due to the nature of the CSMA/CA protocol - especially hampered in high-density scenarios - and the rigidity of the current configurations used by wireless devices (\citealp{akella2007self}), network overlapping drives into many problems and situations that result into poor throughput performance. Our goal is to provide a solution that enhances SR in an online fashion. To this end, we model the problem in which multiple WLANs contend for a common set of resources through adversarial MABs. The adversarial MAB problem (\citealp{auer1995gambling}) frames the scenario in which different learners compete for the same resources simultaneously. In particular, after taking an action, a given learner is granted with a reward that depends on the others' actions, i.e., the joint action profile. 

MABs have been shown to properly deal with the exploration-exploitation trade-off when the uncertainty level is very high (\citealp{auer2002finite, audibert2009exploration, scott2010modern}), which properly addresses the decentralized SR problem studied in this work. Note that learning WLANs do not have information about the environment, which in addition is adversarial because of the competition for the channel resources. Moreover, the interactions shown by CSMA/CA-based devices lead to a very complex problem in terms of finding the optimal solution. Moreover, the lack of data and delay-sensitive constraints prevent using more powerful techniques such as convex optimization or even Deep Learning (DL). These techniques are computationally expensive and require from a lot of training data. In the decentralized SR problem presented here, none of the requirements can be satisfied. To alleviate this, MABs attempt to maximize the achieved performance while the learning procedure is being carried out.

For the remainder of this work, we consider that the concepts of WLAN and agent can be indistinctly exchanged, since WLANs act as learners by collecting knowledge regarding their possible configurations and the experienced throughput. In practice, WLANs accumulate knowledge of a given selected action by observing its performance during a certain amount of time, i.e., a learning iteration. Consequently, the accuracy of long-term estimations depends on for how long the output of a given action is observed. The analysis of the necessary time to successfully monitor the channel is out of the scope of this paper. Thus, we assume perfect long-term estimations regarding the actions' performance. Furthermore, due to the lack of coordination between WLANs, the abovementioned learning procedure would be done in a disorganized way. Accordingly, from a global network perspective, agents would pick actions at any time within a learning iteration, since they are not synchronized in practice. However, and for the sake of simplicity, we consider that WLANs select an action at the beginning of each iteration, so that we can properly capture the performance associated with the different actions (recall that long-term estimates of actions are considered). Therefore, the moment at which adversarial agents select an action is irrelevant to our analysis.

\begin{figure}[h!]
	\centering		
	\begin{subfigure}[b]{0.35\textwidth}
		\includegraphics[width=\textwidth]{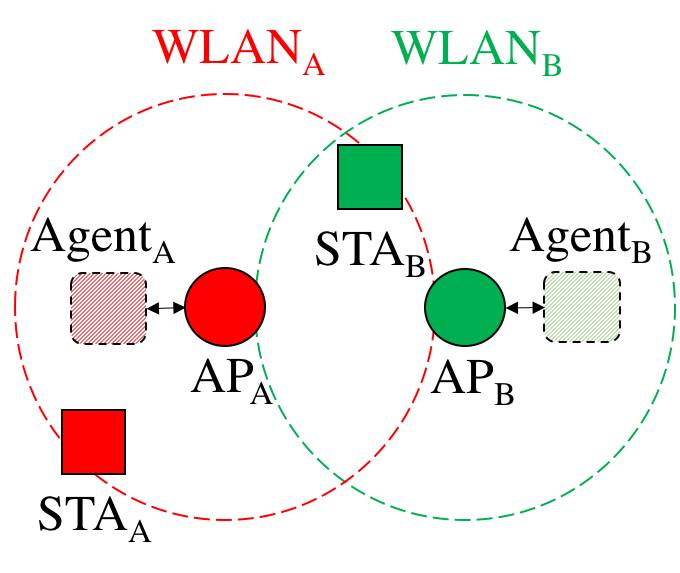}
		\caption{WLANs with agents}\label{fig:agents_a}
	\end{subfigure}
	\begin{subfigure}[b]{0.6\textwidth}
		\includegraphics[width=\textwidth]{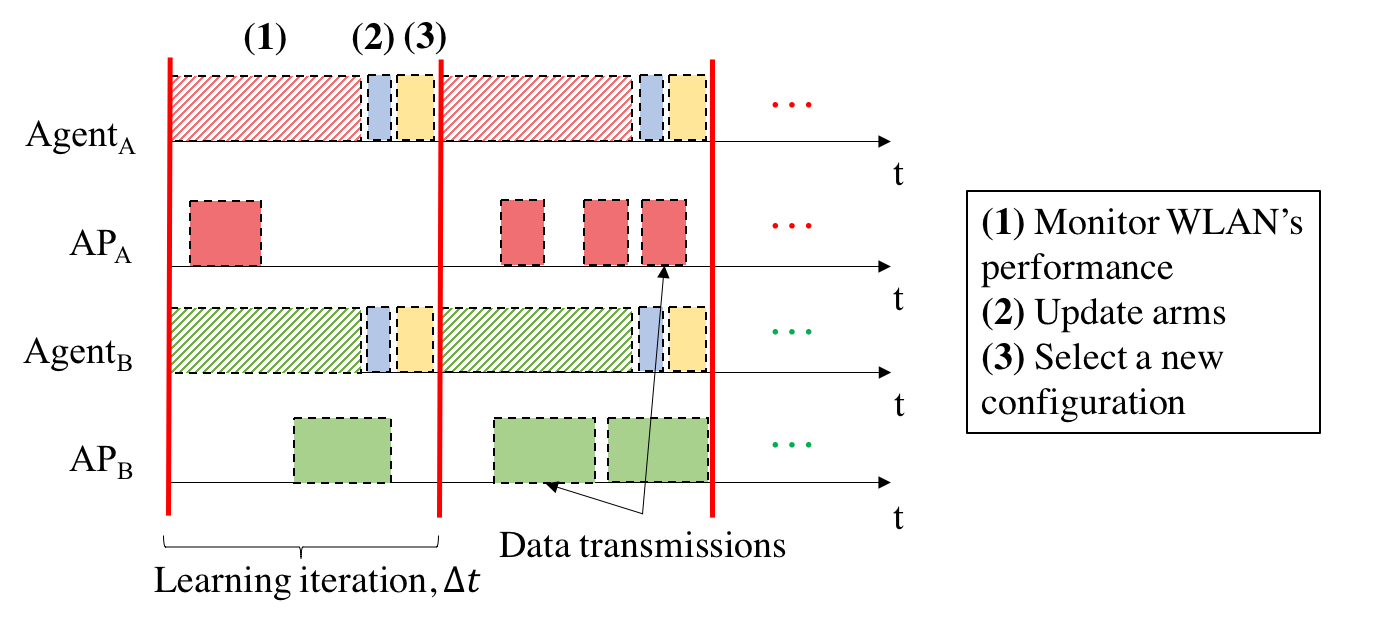}
		\caption{Learning procedure}\label{fig:agents_b}
	\end{subfigure}
	\caption{Agents integration in WLANs. (a) Scenario with two potentially overlapping WLANs, (b) Learning procedure followed by agents according to the performance observed in their associated WLAN.}
	\label{fig:agents}
\end{figure}

Figure \ref{fig:agents} illustrates the inclusion of agents into WLANs, which operate on top of CSMA/CA, as well as the aforementioned learning procedure (Figure \ref{fig:agents_b}). As shown, both agents act within each learning iteration. Initially, an agent observes the performance of the WLAN, which depends on overall network configuration. With such an information, the agent updates the estimate of each action and selects a new one accordingly. This procedure is repeated at the beginning of a new iteration. For the scenario shown in Figure \ref{fig:agents_a}, there is an overlapping between the two WLANs during the initial iteration, and simultaneous transmissions cannot be held. According to this information, a new action is chosen by both Agents A and B, which turns out to enable SR, thus allowing a higher number of successful data transmissions.

Roughly, the SR problem in IEEE 802.11 WLANs can be modeled through adversarial MABs as follows:	
\begin{itemize}
	\item Let $\mathcal{N} = \{1,..., N\}$ be the set of potentially overlapping WLANs.
	\item Each WLAN can choose from a range of actions $\mathcal{A} = \{1,..., K\}$, which refer to combinations of $\mathcal{C}$ non-overlapping frequency channels, $\mathcal{P}$ transmit power levels, and $\mathcal{S}$ sensitivity levels.
	\item Initially, the estimate reward of each action available in any WLAN, $k \in \{1,...,K\}$, is set to 0.
	\item At every iteration, each WLAN selects an arm randomly, according to its action selection-strategy, which in this work is considered to be Thompson sampling.
	\item After choosing an action $k$ at iteration $t$, each WLAN observes the reward generated by the environment, $r_{k,t}$, which is based on the experienced throughput that depends on \emph{i)} its own action and \emph{ii)} the actions made by the overlapping WLANs.
	\item The new information is used for updating the knowledge on the available arms.
\end{itemize}

The goal of an agent, then, is to maximize the reward function, which is equivalent to minimize the accumulated regret. In particular, the accumulated regret $R_{w,T}$ that a given WLAN $w$ experiences until time $T$ can be characterized as follows:
\begin{equation}
R_{w,T} = \sum_{t = 1}^{T} (r_{w,t}^* - r_{w,t}),
\nonumber
\end{equation}
where $r_{w,t}^*$ is the optimal reward granted by the best possible action in iteration $t$, and $r_{w,t}$ is the reward granted by the actual action chosen by WLAN $w$ at that iteration. Since we face an adversarial setting, the process of minimizing the regret is highly influenced by the others' behavior. This raises concerns about the existence of an equilibrium in which the area throughput is fairly maximized. 

For practical application of MABs in WLANs, the reward experienced by a given learner must be normalized, ideally, by the optimal reward $r_{w,t}^*$. This procedure is key to assess the potential of the played actions. But faced by the impossibility of providing such a value for every spatial distribution, which would require an exhaustive search, we define an upper bound consisting in the throughput that a given WLAN obtains in isolation (this concept is further described in Sections \ref{subsubsection:selfish_learning} and \ref{subsubsection:informed_learning}). Finally, it is important to remark that we consider time-invariant rewards, i.e., a given action  always leads to the same reward.

\subsection{Thompson Sampling}
\label{subsection:thompson_sampling}
Thompson sampling has been employed in this work as the action-selection strategy used by WLANs. The motivation behind this choice is that Thompson sampling has been shown to grant excellent performance in front of other well-known policies such as UCB or EXP3, when applied into wireless networks. In (\citealp{wilhelmi2017collaborative}), it was shown to converge fast to the optimal solution in terms of proportional fairness for adversarial environments, thus reducing the temporal variability observed for other exploration-exploitation mechanisms. Essentially, Thompson sampling (\citealp{thompson1933likelihood}) is a Bayesian algorithm that constructs a probabilistic model of the rewards and assumes a prior distribution of the parameters of said model. Given the data collected during the learning procedure, Thompson sampling keeps track of the posterior distribution of the rewards and pulls arms randomly in a way that the drawing probability of each arm matches the probability of the particular arm being optimal. In practice, this is implemented by sampling the parameter corresponding to each arm from the posterior distribution, and pulling the arm yielding the maximal expected reward under the sampled parameter value. 

For the sake of practicality, we apply Thompson Sampling using a Gaussian model for the rewards with a standard Gaussian prior as suggested in (\citealp{agrawal2013further}). By standard calculations, it can be verified that the posterior distribution of the rewards under this model is also Gaussian with mean $\hat{r}_k(t) = \frac{\sum_{w=1:k}^{t-1} r_k(t) }{n_k(t) + 1}$ and variance $\sigma_k^2(t) = \frac{1}{n_k(t) + 1}$, where $n_k(t)$ is the number of times that arm $k$ was drawn until the beginning of round $t$. Henceforth, implementing Thompson sampling in MABs amounts to sampling a parameter $\theta_k$ from the Gaussian distribution $\mathcal{N}\left(\hat{r}_k(t),\sigma_k^2(t)\right)$ and choosing the action $k$ with the highest value. Our implementation of Thompson sampling to the WLAN problem is detailed in Algorithm \ref{alg:thompsons}.	
\begin{algorithm}[h!]
	\SetKwInOut{Input}{Input}
	\SetKwInOut{Output}{Output}		
	Function Thompson Sampling $(\mathcal{A})$\;
	\Input{$\mathcal{A}$: set of possible actions in \{$1, ..., K$\}}
	initialize: $t=0$,  for each arm $k \in \mathcal{A}$, set $\hat{r}_{k} = 0$ and $n_k = 0$ \\
	\While{active}
	{
		For each arm $k \in \mathcal{A}$, sample $\theta_k(t)$ from normal distribution $\mathcal{N}(\hat{r}_{k}, \frac{1}{n_k + 1})$ \\
		Play arm $k = \underset{1,...,K}{\text{argmax }} \theta_k(t) $ \\
		Observe the throughput experienced $\Gamma_t$\\			
		Compute the reward $r_{k,t}$ \\
		$ \hat{r}_{k,t} \leftarrow \frac{\hat{r}_{k,t}  n_{k,t} + r_{k,t}}{n_{k,t} + 2}$\\
		$n_{k,t} \leftarrow n_{k,t} + 1$\\
		$t \leftarrow t + 1$
	}
	\caption{Implementation of MABs (Thompson sampling) in a WLAN}
	\label{alg:thompsons}
\end{algorithm}	

In this paper, the reward is defined in two different ways, which are described in the following subsections.

\subsubsection{Selfish Reward}
\label{subsubsection:selfish_learning}
The first considered reward aims to characterize a selfish behavior, which allows to purely represent the decentralized and adversarial SR problem. Through selfish learning, several WLANs attempt to learn the best configuration for their own gain, regardless of the performance experienced by neighboring networks. In fact, WLANs ignore the existence of other learners, which may have different goals. Henceforth, the reward $r_{w,t}$ that a given learner $w$ experiences at iteration $t$ is computed according to the throughput $\Gamma_{w,t}$ it experiences: 
\begin{equation}
r_{w,t} = \frac{\Gamma_{w,t}}{\Gamma_w^*}, 
\nonumber
\end{equation}		
where $\Gamma_w^*$ is a normalization value that refers to a certain upper bound reward that WLAN $w$ can experience. In the selfish case, the optimal upper bound is given for any configuration that maximizes the individual performance of a given WLAN, regardless of the performance of other WLANs. It is important to remark that it may not be possible for the learner to know such an upper bound (further discussed in Section \ref{subsubsection:reward}). In consequence, and for the rest of this paper, we define the upper bound reward to be the throughput that a given WLAN would obtain in isolation.

Selfish learning in WLANs has been shown to potentially increase SR while leading to collaborative results, provided that competitors enjoy equal opportunities (\citealp{wilhelmi2017collaborative}). However, unfairness issues may be unleashed when dealing with significant asymmetries in terms of nodes location. As a result, WLANs in a dominant position may learn a performance maximization strategy at the expense of harming the weaker ones. By extension, competition among nodes is prone to lead to suboptimal configurations, so that the optimal action ends up being hidden to learners. In this sense, from the learner's point of view, the right action may not be robust enough against the environment, as a result of being susceptible to outer aggressive actions. Furthermore, learning selfishly in an adversarial environment may be detrimental in terms of temporal throughput variability. 

\subsubsection{Environment-Aware Reward}
\label{subsubsection:informed_learning}
To overcome the unfairness situations that may be generated by selfish learning, we propose the environment-aware reward, which takes into consideration the effects that the actions of a given learner have on the environment (i.e., on the overlapping WLANs). To this end, we assume that WLANs are able to estimate the others' performance by listening to their activity on the channel. In practice, estimating the throughput experienced by overlapping WLANs may have limitations and lead to inaccurate values. Nevertheless, we assume perfect estimation to purely study the benefits and drawbacks of environment-aware learning. The analysis of dealing with inaccurate estimations is left as future work.

By assuming the availability of environmental information, we define an environment-aware reward that aims to fairly enhance the area throughput. Henceforth, rather than letting WLANs use their own performance, we propose that the reward experienced by each WLAN includes some notion of fairness. Three well-known fairness metrics are: \emph{i)} Jain's Fairness Index (JFI) of the throughput, \emph{ii)} Proportional Fairness (PF) of the throughput, and \emph{iii)} max-min throughput. Throughout this paper, we are considering only the latter, since the JFI does not necessarily maximize aggregate performance, and the PF is very varying.\footnote{Very different results may lead to the same (or very similar) PF value, which may have consequences on the learning procedure followed by WLANs. For instance, regarding the performance of two WLANs, a completely fair distribution of [50, 50] Mbps leads to a similar PF than a much more unfair distribution of [120, 20] Mbps.} As a result, the reward $r_{\mathcal{O},t}$ that a set of $\mathcal{O}$ overlapping networks experience in iteration $t$ is given by:
\begin{equation}
	r_{\mathcal{O},t} =\frac{\min_{w \in \mathcal{O}} \Gamma_{w,t}}{\Gamma_{\mathcal{O},t}^*}, 
	\nonumber
\end{equation}			
where $\min_{w \in \mathcal{O}} \Gamma_{w,t}$ is the minimum throughput experienced in the set of overlapping WLANs $\mathcal{O}$. The upper bound reward $\Gamma_{\mathcal{O},t}^*$ is shared, and refers to the configuration that grants the maximum max-min throughput. Again, since this knowledge may not be known at the learner side, we consider the set of throughputs in isolation for each WLAN in $\mathcal{O}$. Then, the max-min throughput value is taken as the shared optimal reward.

\subsection{Considerations of Decentralized Learning in WLANs}
\label{subsection:considerations}
The classical MAB problem frames the scenario whereby an agent interacts with the environment. The agent's goal is to maximize the long-term reward according to the actions it plays, regardless of any external factor. However, the presence of other agents in the adversarial learning problem adds an extra layer of complexity. That is the case of decentralized SR, where different potentially overlapping WLANs aim to find the best configuration by their own. 

The competition unleashed by the adversarial setting can be formulated from a game theoretical perspective. It is important to concentrate on the possible equilibriums that can be achieved for a given game, which can be defined by the set of competitors and their strategies. Of course, reaching an equilibrium whereby performance is maximized is limited to the conflicts that may crop up as a result of the clashing strategies followed by different players. For instance, the performance of a set of overlapping WLANs can be significantly limited if they use aggressive strategies in terms of interference generation, since a suboptimal equilibrium may be reached. In case of using the selfish reward presented in this paper, aggressive strategies in terms of interference would be preferred by learners, specially in dense environments. Note that WLANs seek to maximize their individual throughput, regardless the performance of the other networks. This particular scenario is further analyzed in Section \ref{subsubsection:competition}. Moreover, it is possible that an equilibrium cannot be found in a decentralized manner. The main reason lies in the scarcity of the resources being shared, and in the individual requirements of each WLAN. In that case, if greedy strategies were employed, WLANs would alternate good and bad performing actions. 
	
As a consequence to the adversarial setting unleashed in the decentralized SR problem, some important implications must be considered with regards to practical application of MABs to WLANs. Note, as well, that even if using an environment-aware reward that promotes collaboration (WLANs share a common goal), learning in a decentralized way may result into some other performance limitations. Such an issue is studied in Section \ref{subsubsection:clustering}. In essence, implications are noticed on the action-selection procedure, i.e., the set of rules and constraints according to which a given agent learns from the environment. Such a followed procedure is key to determine the potential of a given algorithm in terms of achievable performance and convergence guarantees. In the decentralized SR problem, the action selection procedure is held in a disorganized way, since every agent attempts to learn by its own. Such a situation leads to highly-varying environments in which an intensive action-selection procedure is held. This may severely impact on the learning process followed by any learner, and is worsened as the number of overlapping learners increases.

Regarding the learning process, on the one hand, a sublinear regret cannot always be guaranteed because of the intensive competition among networks. The speed at which regret is minimized strongly depends on the scenario. Because of the adversarial setting, a zero-regret configuration may be not be found, even if it exists. As a direct consequence, learners may suffer an increased variability on the experienced reward. Such a statement differs from the current work in multi-player MABs for opportunistic spectrum access, where strategies can be defined for sublinear regret minimization. First of all, unlike the SR problem, actions' performance can be binary modeled when attempting to access the channel, thus allowing to extract much more meaningful information regarding the environment: if selecting a given channel leads to a high number of collisions, the learner can easily infer that it is saturated. In the SR case, however, much more complex interactions may occur and have implications on a per-WLAN basis. Then, it is the aim of this work to provide insights on the application of decentralized learning to maximize the performance of a wireless network.

Finally, and related to regret minimization, assessing convergence in a WLAN (i.e., stop acting) may not be possible for the SR problem, thus impacting on the performance of higher communication layers. In particular, we can determine that a WLAN has learned which is the optimal action if it experiences a regret below a given threshold. However, due to the adversarial setting, such a condition may not hold, or may not be accomplished before the environment changes. 

\subsubsection{Reward Definition}
\label{subsubsection:reward}
A reward function describes how an agent should ideally behave, which allows conducting its activity towards maximizing (or minimizing) a given performance metric, i.e., the learner shapes a policy according to the obtained rewards. By extension, a precise definition of the reward allows to improve the learning procedure, since the reward perfectly matches with the desired goal. In that case, convergence can be improved, and the probability of falling into a local minimum is lower. 

Unfortunately, defining a reward function in practice may become a very complex task. On the one hand, the optimal performance that can be achieved by a given individual or set of agents may not be known, thus hindering the learning procedure. On the other hand, the reward-based policies constructed by agents may be limited because of the competition among nodes (either for the selfish or the environment-aware approaches). Such dependencies can result in dominance positions of certain policies above others, which may obfuscate the optimal solution (obtained by the non-dominant policies). Moreover, reaching the optimal behavior is subject to the convexity of the joint reward function, which is not the case for the presented SR problem.

\begin{figure}[h!]
	\centering
	\begin{subfigure}[b]{0.25\textwidth}
		\includegraphics[width=\textwidth]{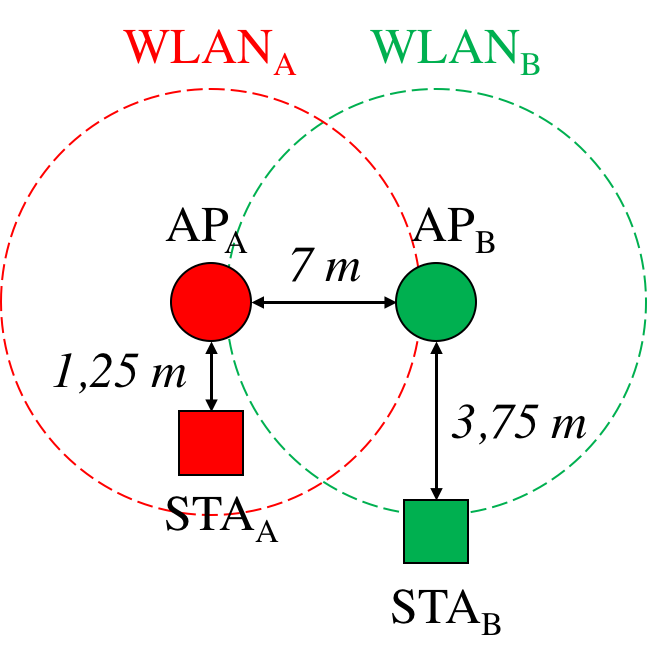}
		\caption{Scenario}
		\label{fig:selfish_s3}
	\end{subfigure}
	\begin{subfigure}[b]{0.3\textwidth}
		\includegraphics[width=\textwidth]{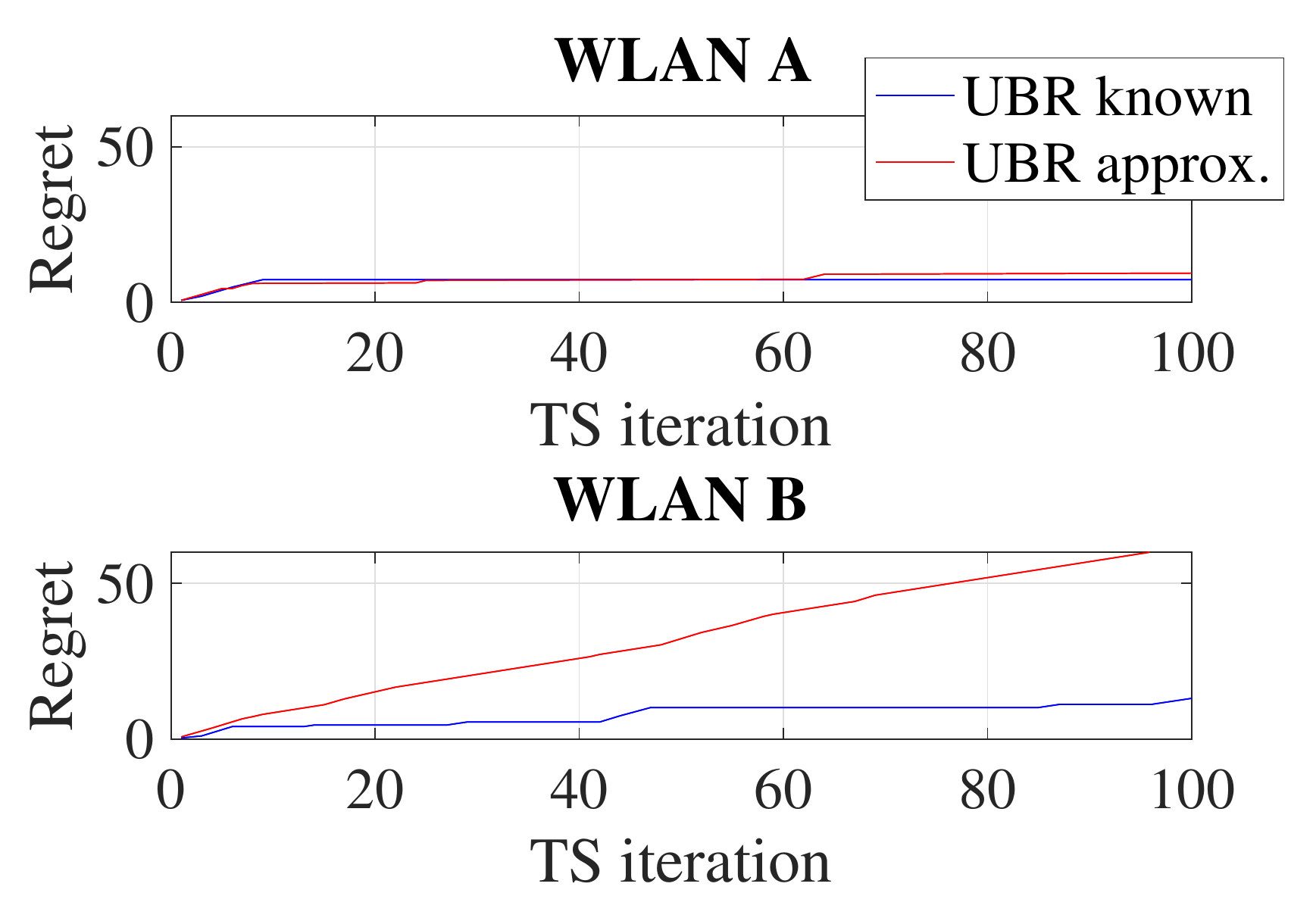}
		\caption{Average regret}
		\label{fig:approx_vs_actual_regret}
	\end{subfigure}
	\begin{subfigure}[b]{0.3\textwidth}
		\includegraphics[width=\textwidth]{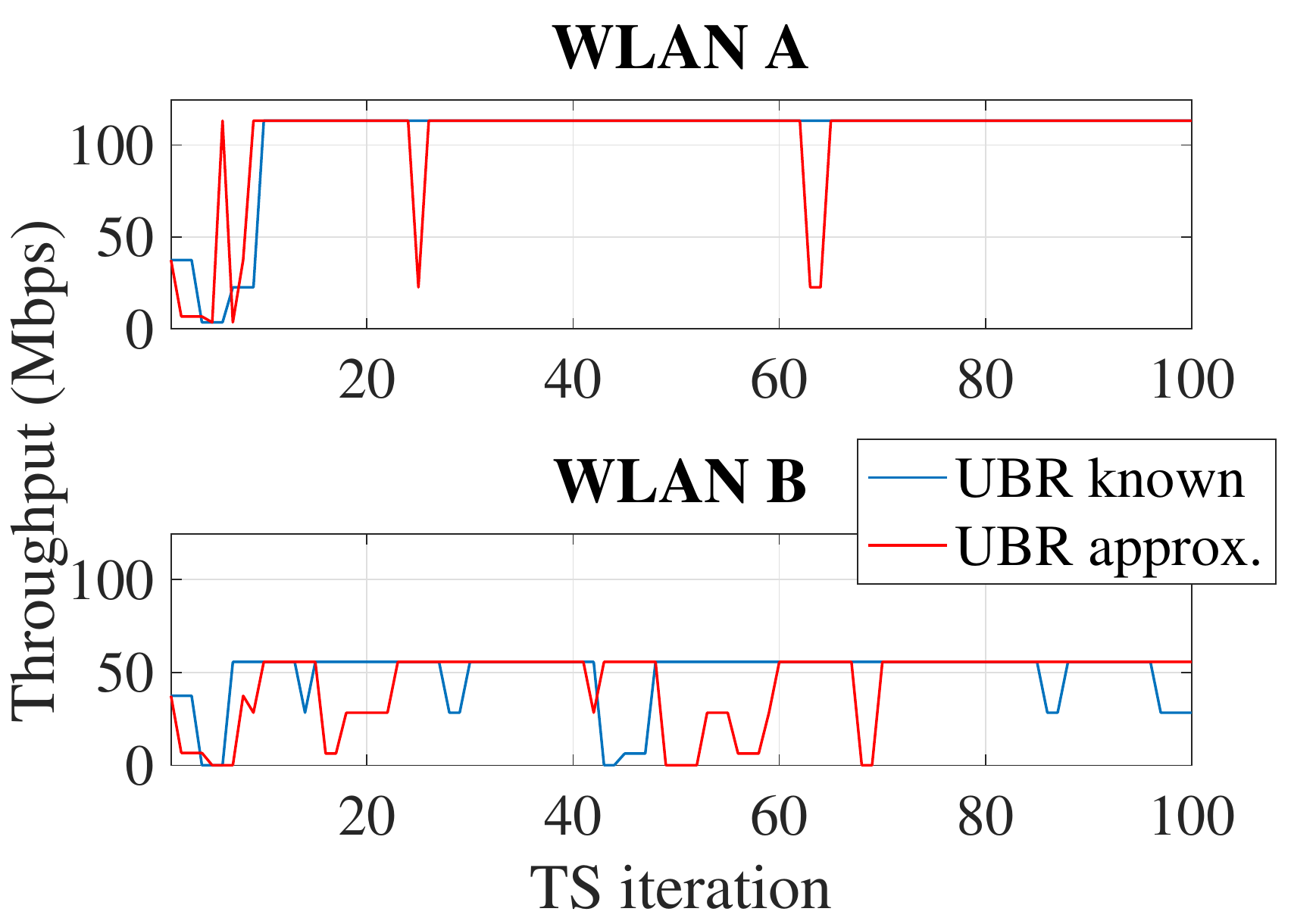}
		\caption{Temporal throughput}
		\label{fig:approx_vs_actual_tpt}
	\end{subfigure}
	\caption{Upper bound reward considerations when applying selfish Thompson sampling (100 iterations are considered). (a) Scenario with two asymmetric WLANs in terms of maximum capacity, (b) Temporal regret experienced by each WLAN when the actual upper bound reward (UBR) is known (blue) or not (red), (c) Temporal throughput experienced by each WLAN when the actual UBR is known (blue) or not (red).}	
	\label{fig:actual_vs_approximated_reward}
\end{figure}

Now, in order to illustrate the impact of approximating the reward in the SR problem, let us consider a simple scenario (depicted in Figure \ref{fig:selfish_s3}) and focus only on the selfish reward type (previously defined in Section \ref{subsubsection:selfish_learning}). In particular, we place two WLANs that apply Thompson sampling selfishly. The range of possible actions in terms of CCA and transmit power levels are defined in Table \ref{tbl:simulation_parameters} (included in \ref{section:simulated_wireless_environment}). We compare the usage of a reasonable upper bound reward (given a decentralized environment) in front of the optimal performance that can be actually achieved (computed by brute force). For the former, based on IEEE 802.11ax PHY capabilities (refer to simulation parameters in \ref{section:simulated_wireless_environment}), we use the theoretical data rate that can be achieved in case of using the maximum MCS.\footnote{We assume that the maximum data rate is achieved in case of using a single user (SU) transmission through a 1024-QAM MCS and a coding rate of 5/6. According to the IEEE 802.11ax standard parameters, this leads to a data rate of 114.37 Mbps.} Note, as well, that this data rate may not correspond to the actual optimal performance due to several factors such as nodes position and inter-WLAN interactions. However, we refer to the utilization of a fixed MCS as an illustrative example of a practical upper bound that could be used in real networks. In contrast, we will use the throughput in isolation as an upper bound later in Section \ref{section:performance_evaluation}.

Figures \ref{fig:approx_vs_actual_regret} and \ref{fig:approx_vs_actual_tpt} show the experienced regret and throughput, respectively, experienced by two overlapping WLANs when applying Thompson sampling selfishly during 100 iterations. In order to emphasize on the effects of using an inaccurate upper bound, one of the WLANs (namely, $\text{WLAN}_\text{B}$) has stronger limitations than the other one (namely, $\text{WLAN}_\text{A}$), whose AP-STA distance is shorter. Such a situation makes $\text{WLAN}_\text{B}$ more vulnerable in front of interference and prevents it to achieve the highest achievable throughput due to the SINR sensed at the receiver. As shown in Figure \ref{fig:approx_vs_actual_tpt}, $\text{WLAN}_\text{B}$ experiences a higher throughput variability in case of using an approximated upper bound reward, rather than using the actual information for this concrete scenario. This can be also noticed in Figure \ref{fig:approx_vs_actual_regret}, where the regret experienced by $\text{WLAN}_\text{B}$ grows linearly if the actual optimal performance is unknown. In contrast, $\text{WLAN}_\text{A}$ is able to use the maximum MCS due to its privileged situation, thus showing similar performance both for known and approximated upper bounds.

When defining an upper bound reward, we have seen that false expectations may lead to non-convergence, which may have severe implications in the temporal variability of the experienced performance.

\subsubsection{Neighbors Identification when Applying an Environment-Aware Reward}
\label{subsubsection:clustering}
In environment-aware learning, WLANs take the others' performance into account during the reward generation process. However, estimating the throughput of neighboring WLANs raises the following question: which are the potentially overlapping WLANs that each learner should consider? The fact of dealing with complex spatially-distributed environments hinders answering to that question, since interactions in overlapping WLANs are not trivial to be derived for the SR problem, and change with time. As a result, for a given learner, it is hard to identify the set of potentially overlapping WLANs whose performance must be taken into account. For convenience, let us refer to a particular set of overlapping networks as a cluster. In accordance to that, a WLAN applying clustering refers to the procedure whereby it considers the performance of other overlapping networks during the reward generation process. Note, as well, that an overlap between two networks may occur by different reasons. For instance, one may consider that two WLANs overlap if the mutual generated interference exceeds a given threshold, which may not necessarily be the capture threshold. For the rest of this paper, we assume that WLANs sharing a reward only take consideration of those generating a level of interference greater than the CCA threshold on their own. Furthermore, we assume bidirectional interactions, even in presence of asymmetries.

To showcase the importance of properly defining a list of neighbors (i.e., clusters), let us define a simple scenario in which two WLANs are independent to one another in terms of interference. Such a scenario has the particularity that one WLAN has limited performance due to the AP-STA distance. Therefore, we aim to study the effects of learning by either considering all the environment (long-range cluster) or just the interfering devices (short-range cluster). On the one hand, we establish a soft establishment rule, where a neighbor is considered if the received power is higher than a very low decision threshold. In practice, this is equivalent to not considering any neighbors establishment rule, so that the max-min fairness involves all the WLANs in the presented scenario. On the other hand, short-range clustering by SINR is done (previously introduced in Section \ref{subsubsection:informed_learning}), which means that the performance of a given WLAN is considered by another one if the power received from the former is greater than the latter's CCA threshold.

\begin{figure}[h!]
	\centering   		
	\begin{subfigure}[b]{0.35\textwidth}
		\includegraphics[width=\textwidth]{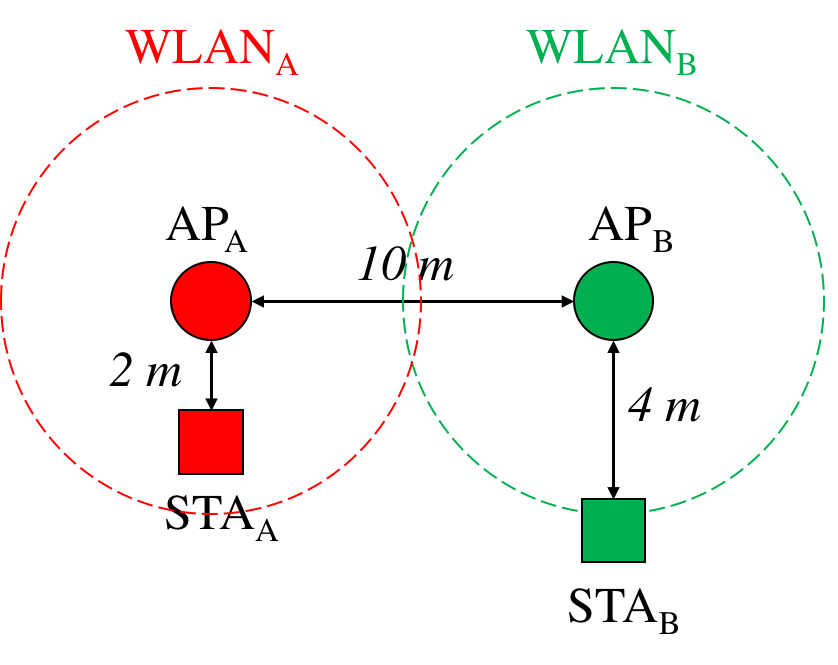}
		\caption{Scenario}
		\label{fig:informed_s1}
	\end{subfigure}
	\begin{subfigure}[b]{0.4\textwidth}
		\includegraphics[width=\textwidth]{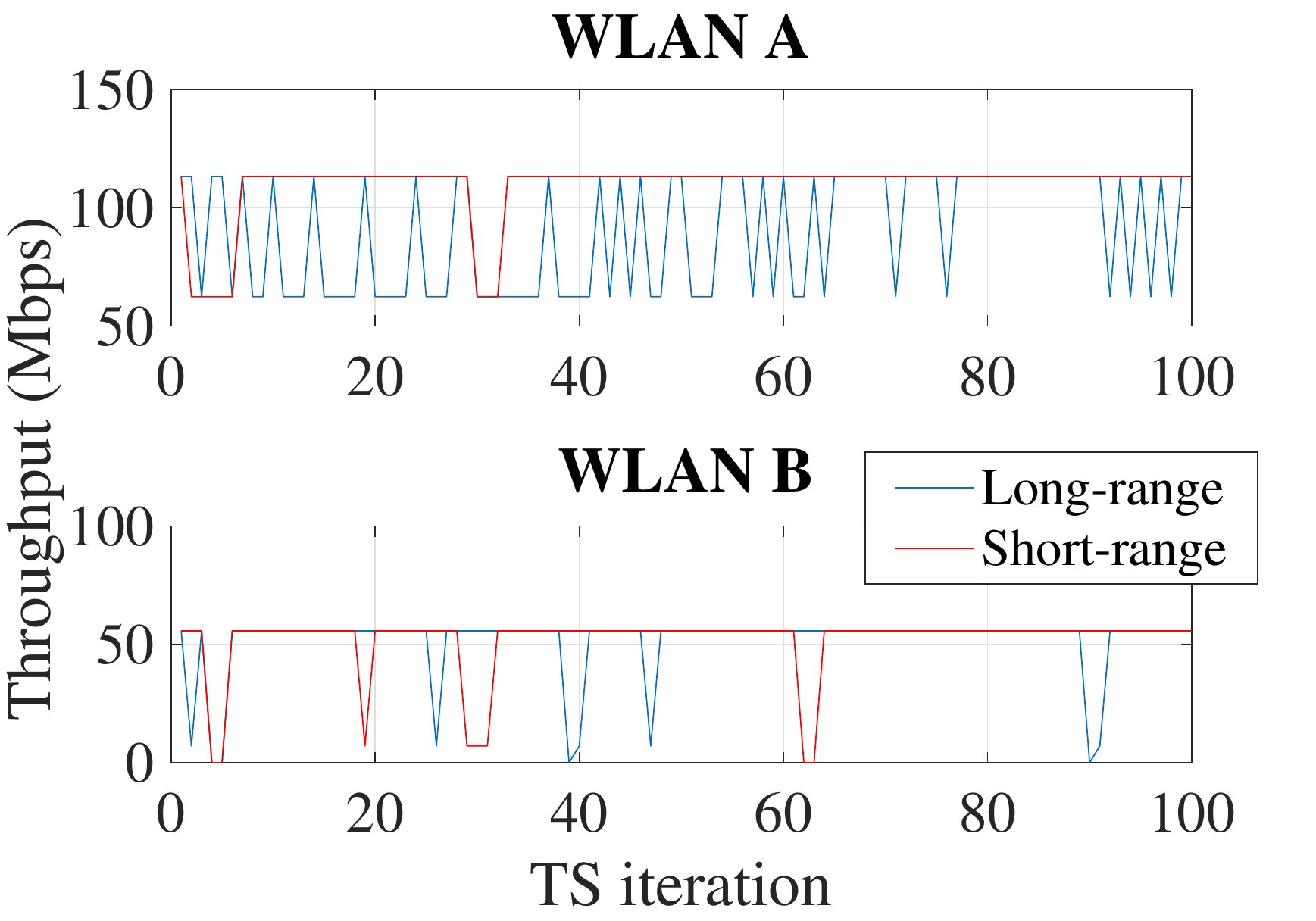}
		\caption{Individual throughput}
		\label{fig:clustering_benefits}
	\end{subfigure}
	\caption{Neighbors establishment considerations when applying environment-aware Thompson sampling (100 iterations are considered). (a) Scenario with two independent WLANs in terms of interference, (b) Temporal throughput experienced by each WLAN when using long-range (blue) and short-range clustering (red).}
	\label{fig:clustering}
\end{figure}   	

As shown in Figure \ref{fig:clustering_benefits}, short-range clustering grants better results in terms of temporal variability, since $\text{WLAN}_\text{A}$ does not consider $\text{WLAN}_\text{B}$ during the learning procedure (the CCA condition does not hold). Otherwise, in case that long-range clustering is applied, $\text{WLAN}_\text{A}$ cannot determine that $\text{WLAN}_\text{B}$ is not a potential overlapping network, i.e., the actions of the latter do not impact to performance of the former. If long-range clustering is applied, $\text{WLAN}_\text{A}$ considers the throughput of $\text{WLAN}_\text{B}$ to learn the performance of each action. This prevents the former to distinguish which are the best actions for itself. Therefore, good and bad performing actions are alternated because of the capacity limitation of $\text{WLAN}_\text{B}$ (it never becomes satisfied).

\begin{figure}[h!]
	\centering   		
	\begin{subfigure}[b]{0.35\textwidth}
		\includegraphics[width=\textwidth]{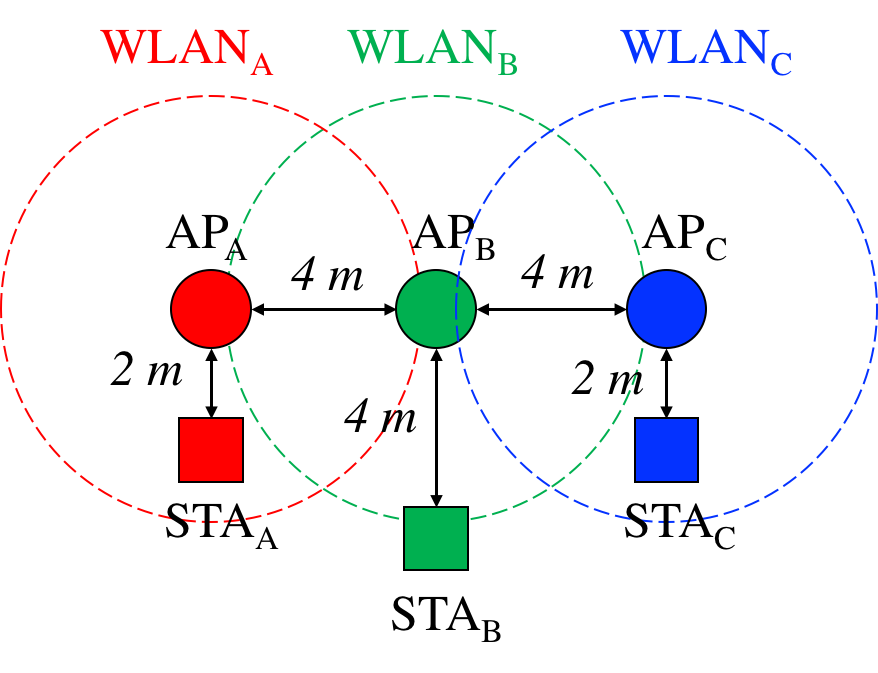}
		\caption{Scenario}
		\label{fig:informed_s2}
	\end{subfigure}
	\begin{subfigure}[b]{0.4\textwidth}
		\includegraphics[width=\textwidth]{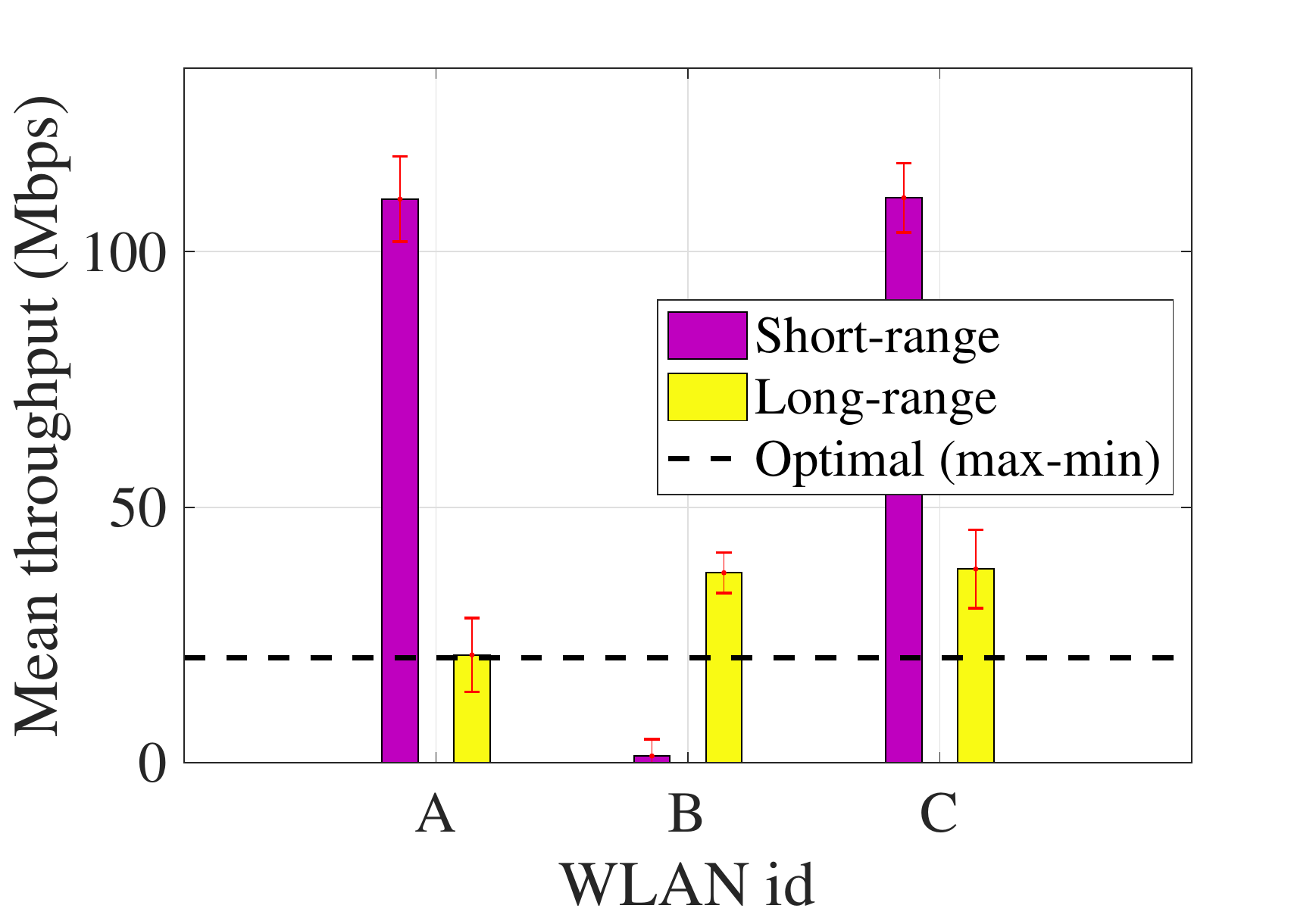}
		\caption{Average throughput}
		\label{fig:clustering_additive_interference}
	\end{subfigure}
	\caption{Issues on neighbors establishment when applying environment-aware Thompson sampling (1,000 iterations are considered). (a) Scenario in which $\text{WLAN}_\text{B}$ is prone to suffer from flow starvation, (b) Average throughput per WLAN when using long-range (yellow) and short-range clustering (purple). The standard deviation of the average throughput between iterations is shown in red, and the black dashed line indicates the shared goal.}
	\label{fig:clustering_issues}
\end{figure}

Despite of the remarkable benefits of short-range clustering-based methods, determining neighbors lists is not trivial in dense WLAN scenarios. In particular, the proposed approach in which the SINR is used to determine interfering nodes fails in capturing additive interference situations. Such kind of interference appears when a given network is only affected when two or more WLANs transmit simultaneously. To illustrate this concept, we focus on the scenario shown in Figure \ref{fig:informed_s2}, where additive interference generates flow-in-the-middle starvation to a WLAN located in the middle of the other two. In Figure \ref{fig:clustering_additive_interference} we show the results of applying environment-aware Thompson sampling for both short-range and long-range clustering. This time we have considered the results after 1,000 learning iterations, since we are interested in showing the long-term performance achieved in both situations.

As shown in Figure \ref{fig:clustering_issues}, the short-range clustering approach fails because additive interference affecting $\text{WLAN}_\text{B}$ cannot be captured by the set of overlapping WLANs. Note, as well, that $\text{WLAN}_\text{A}$ and $\text{WLAN}_\text{C}$ always sense the channel free, regardless of the networks that are currently transmitting. This allows them to experience the highest possible throughput. In contrast, when agents consider all the WLANs in the environment (long-range clustering), the starvation at $\text{WLAN}_\text{B}$ is noticed by the others. As a result, collaboration is enabled and the max-min throughput is maximized at the expense of the aggregate performance.

\subsubsection{Learning in Dynamic WLANs}

\begin{figure}[h!]
	\centering   		
	\includegraphics[width=0.4\textwidth]{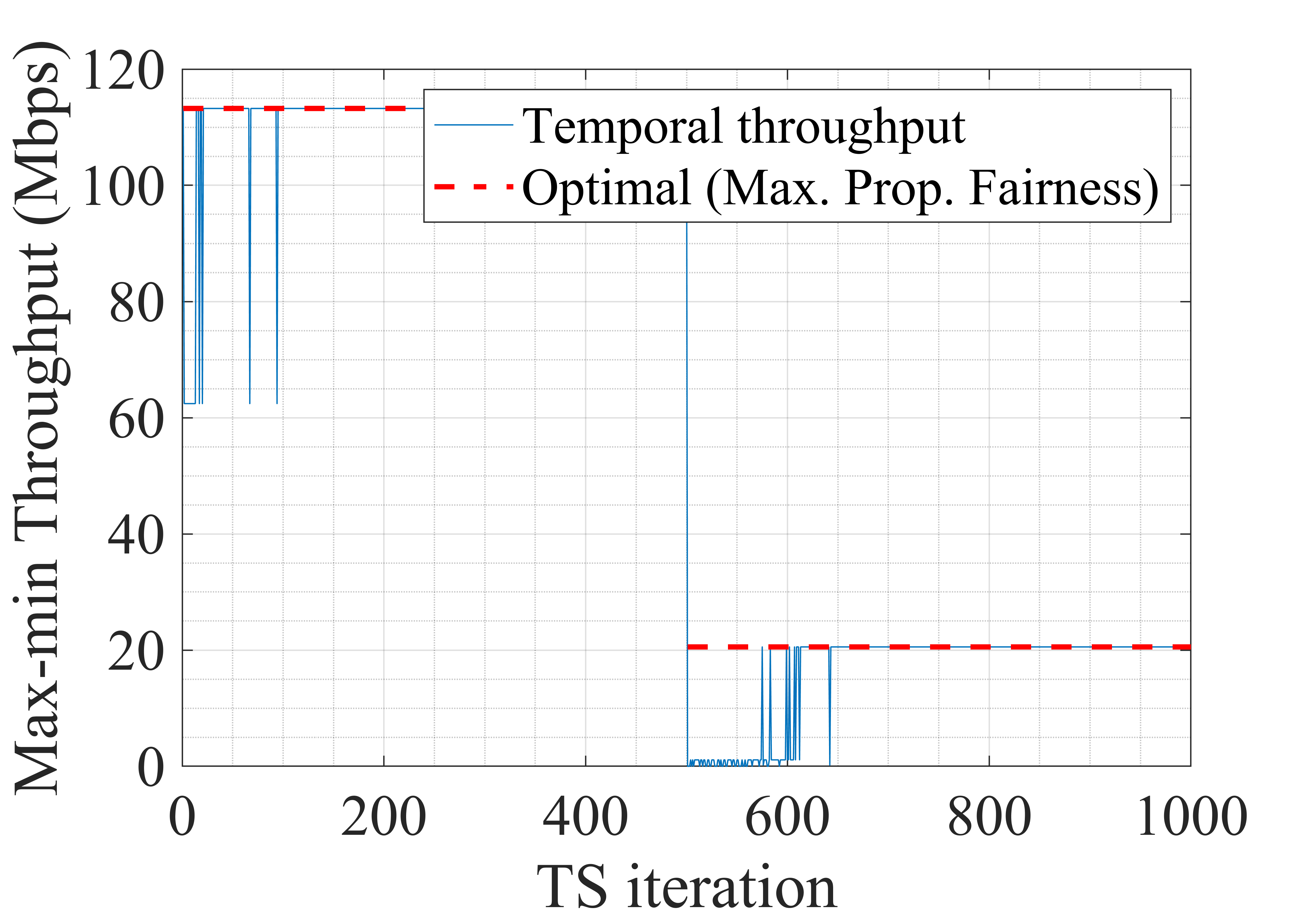}
	\caption{Thompson sampling application in a dynamic scenario where $\text{WLAN}_\text{B}$ appears in iteration 500 (1,000 iterations are considered). The black dashed line indicates the shared goal.}
	\label{fig:dynamic_wlan}
\end{figure}

In addition to the abovementioned learning implications, it is interesting to study the effects of applying MABs in dynamic wireless environments. To that purpose, we frame a scenario in which a new WLAN appears after some other WLANs have already learned the optimal configuration. In particular, we use the scenario shown in Figure \ref{fig:informed_s2}, and consider that $\text{WLAN}_\text{B}$ is activated half-way through the simulation. Figure \ref{fig:dynamic_wlan} shows the max-min throughput achieved when all the WLANs apply Thompson sampling (the environment-aware method is considered). $\text{WLAN}_\text{B}$ is activated at iteration 500, point at which it is expected that $\text{WLAN}_\text{A}$ and $\text{WLAN}_\text{C}$ have gathered enough information for maximizing their performance during the initial phase.

As shown in Figure \ref{fig:dynamic_wlan}, when $\text{WLAN}_\text{B}$ is activated, Thompson sampling adapts to the new situation. It reaches the new optimal goal after a reasonable number of iterations (similar to the initial learning phase). In particular, Thompson sampling needs some time to reshape the already defined probability distributions. The time the algorithm takes to adapt itself can be enhanced if changes in the environment are tracked (e.g., by sensing beacons from new WLANs). Hence, additional information provided to the learning algorithm may boost the exploration of the new optimal actions. In contrast, the implications in terms of performance provoked by changes in the environment are hard to track in practice. Moreover, the procedure to be followed when a change is detected at the algorithm level is not easy to derive. There exists a trade-off between the past and the present information, which must be carefully balanced to achieve optimal performance.

In summary, learning in dynamic environments raises several questions about the validity and expiration of the learned data. This analysis is out of the scope of this paper, so we leave it as future work. In anticipation of this research topic, we highlight that dynamic MABs have been previously studied in (\citealp{hartland2006multi}). Also, with a higher degree of relation to the algorithms shown in this work, we find Dynamic Thompson Sampling (DTS) (\citealp{gupta2011thompson}), which is shown to adapt faster to changes in the environment than the traditional Thompson sampling. Roughly, DTS promotes adaptive exploration by tracking the reward probabilities of each arm, which is useful to give emphasis to recent observations.

\section{Performance Evaluation}			
\label{section:performance_evaluation}				
In this Section we evaluate both selfish and environment-aware decentralized learning strategies. To that purpose, we first study the 
behavior shown by WLANs in representative scenarios when applying both kinds of learning. Then, we generalize those results through simulations in random high-density scenarios.

\subsection{Selfish vs Environment-Aware Learning}
\label{subsection:selfish_vs_informed}  
Selfish and environment-aware strategies are now evaluated in scenarios describing different casuistic. In order to assess the performance achieved in WLANs by applying each strategy, we define the optimal result as: \emph{i)} the maximum individual throughput that a given WLAN can achieve (regardless of the others' performance), \emph{ii)} the maximum throughput that each WLAN can achieve by ensuring the max-min principle. Note, as well, that such values are computed by brute force in the following illustrative scenarios. It is also worth noting that such an optimal performance may not be achieved due to the interactions between WLANs, but special attention will be given to the behavior of each learning approach in relation to that.

For the sake of highlighting Thompson sampling performance in front of other online learning techniques, in this subsection we also provide the results of applying the $\varepsilon$-greedy action-selection strategy. In contrast to Thompson sampling, $\varepsilon$-greedy selects the action with highest absolute performance with probability $1-\varepsilon$ (exploitation), where $\varepsilon$ is within 0 and 1. Otherwise, a random action is selected with probability $\varepsilon$ (exploration). Note, as well, that the parameter $\varepsilon$ is initialized to 1 and dynamically adjusted as a function of the number of iterations, as done in (\citealp{auer2002finite}).

\subsubsection{Learning in Presence of Asymmetries}
\label{subsubsection:fairness}  	
Wireless networks are not always symmetric in terms of nodes location, so that different WLANs may not enjoy the same opportunities when tackling the environment. Such an issue is more common to occur in dense environments where the diversity of deployments is high. In these situations, attempting to maximize spectral efficiency in a selfish way can be detrimental in terms of fairness, especially if there are WLANs in worse conditions than others. Conversely, the environment-aware approach is expected to solve the imbalance between WLANs by maximizing the max-min throughput.

\begin{figure}[h!]
	\centering   		
	\begin{subfigure}[b]{0.26\textwidth}
		\includegraphics[width=\textwidth]{s1_new}
		\caption{Scenario}
		\label{fig:s1_new}
	\end{subfigure}
	\begin{subfigure}[b]{0.36\linewidth}
		\includegraphics[width=\textwidth]{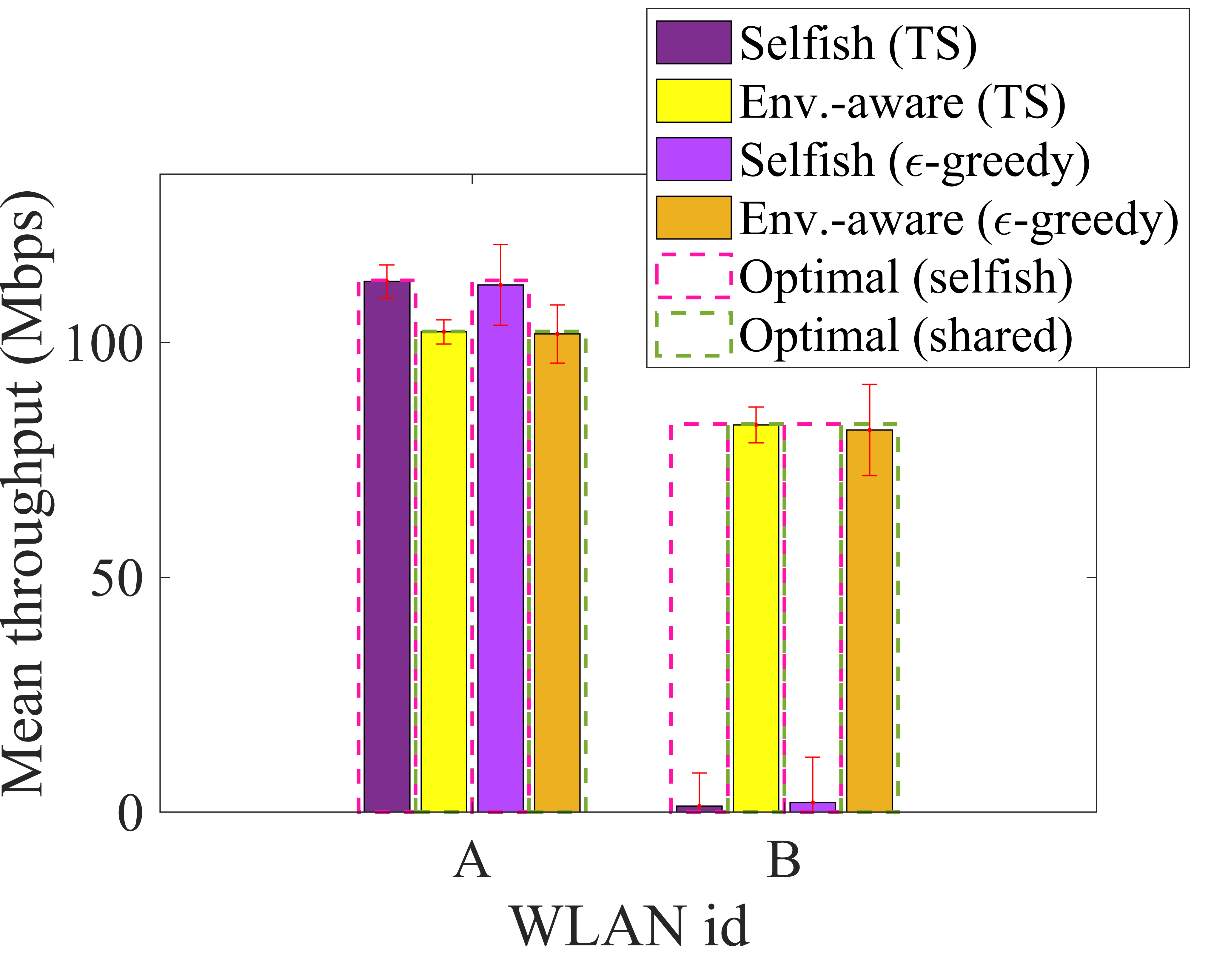}
		\caption{Mean throughput}\label{fig:scenario_1_new}
	\end{subfigure}
	\begin{subfigure}[b]{0.36\textwidth}
		\includegraphics[width=\textwidth]{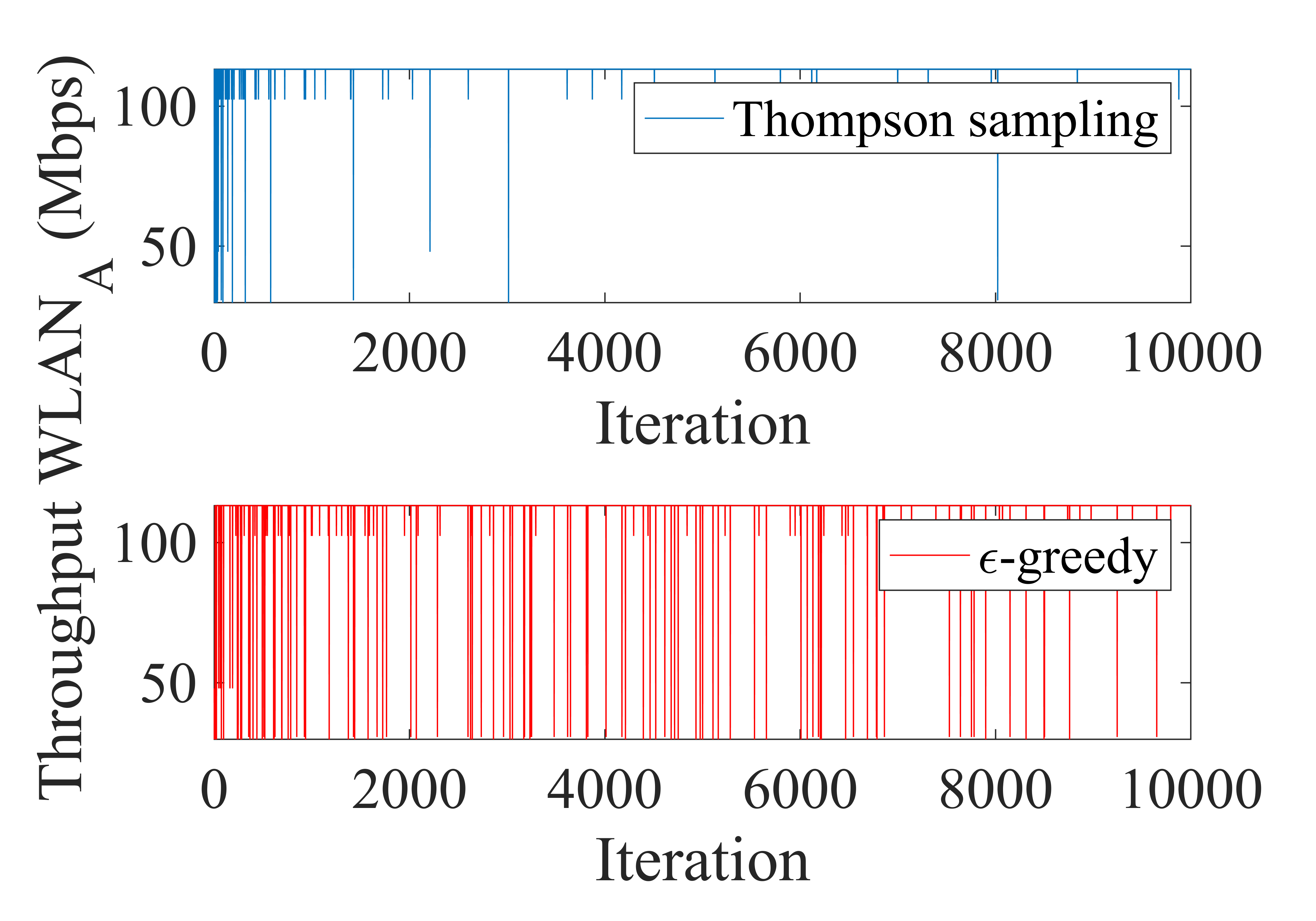}
		\caption{Throughput $\text{WLAN}_\text{A}$}
		\label{fig:experiment_2_1_variability}
	\end{subfigure}
	\caption{Fairness issues in both selfish and environment-aware Thompson sampling (10,000 iterations are considered). (a) Scenario in which $\text{WLAN}_\text{B}$ is prone to suffer from starvation, (b) Average throughput per WLAN for selfish and environment-aware learning, both for Thompson sampling (TS) and $\varepsilon$-greedy. The standard deviation of the average throughput between iterations is shown in red. The pink and green dashed lines indicate the maximum performance achieved per WLAN regarding both selfish and environment-aware optimal solutions, respectively. (c) Temporal throughput achieved by $\text{WLAN}_\text{A}$ when learning selfishly through Thompson sampling and $\varepsilon$-greedy, respectively.}
	\label{fig:selfish_learning_fairness_issue}
\end{figure}

To illustrate the effect of applying both selfish and environment-aware rewards in an asymmetric network, let us retrieve the simple 2-WLANs asymmetric scenario used in Section \ref{subsubsection:reward} (now shown in Figure \ref{fig:s1_new}). In this scenario, each AP is separated $d_{\text{AP}_\text{A},\text{AP}_\text{B}}$ meters from the other one. The distance between an AP and its associated STA is $d_{\text{AP}_\text{A},\text{STA}_\text{A}}$ and $d_{\text{AP}_\text{B},\text{STA}_\text{B}}$, respectively, so that $d_{\text{AP}_\text{A},\text{AP}_\text{B}} > d_{\text{AP}_B,\text{STA}_\text{B}} > d_{\text{AP}_\text{A},\text{STA}_\text{A}}$. 

The results of applying both Thompson sampling and $\varepsilon$-greedy methods for 10,000 iterations are shown in Figures \ref{fig:scenario_1_new} and \ref{fig:experiment_2_1_variability}. While the former indicates the average throughput experienced by each WLAN, the latter illustrates the temporal variability at $\text{WLAN}_\text{A}$.\footnote{We show the performance of only one WLAN as it is representative of the effects we illustrate here.}

As Figure \ref{fig:scenario_1_new} reveals, learning selfishly allows $\text{WLAN}_\text{A}$ to experience the highest possible throughput, both using Thompson sampling and $\varepsilon$-greedy. However, $\text{WLAN}_\text{B}$ suffers from starvation because none of its possible actions allows to palliate the effects of $\text{WLAN}_\text{A}$'s aggressive configuration. In contrast, when both WLANs use the environment-aware strategy, the optimal max-min throughput is achieved, so that the starvation problem in $\text{WLAN}_\text{B}$ is solved. In exchange, $\text{WLAN}_\text{A}$ sacrifices a portion of its maximum achievable throughput, since it uses a less aggressive configuration. The difference between Thompson sampling and $\varepsilon$-greedy lies in the experienced temporal throughput variability, which is significantly higher for the latter method (refer to Figure \ref{fig:experiment_2_1_variability}). Such a variability entails that WLANs experience a slightly lower mean throughput.

As shown in this subsection, selfish learning is prone to generate flow starvation. However, it is worth noting that it is a very common situation in real dense deployments, even if configurations remain static.

\subsubsection{Learning on Equal Terms}
\label{subsubsection:resources_maximization}

We previously analyzed the effect of applying selfish and environment-aware learning in an asymmetric deployment. However, that might not represent other topologies where competing WLANs are in similar conditions. Therefore, we now showcase the potential of applying RL in dense scenarios where WLANs can access to the channel on equal terms. 

\begin{figure}[h!]
	\centering
	\begin{subfigure}[b]{0.26\textwidth}
		\includegraphics[width=\textwidth]{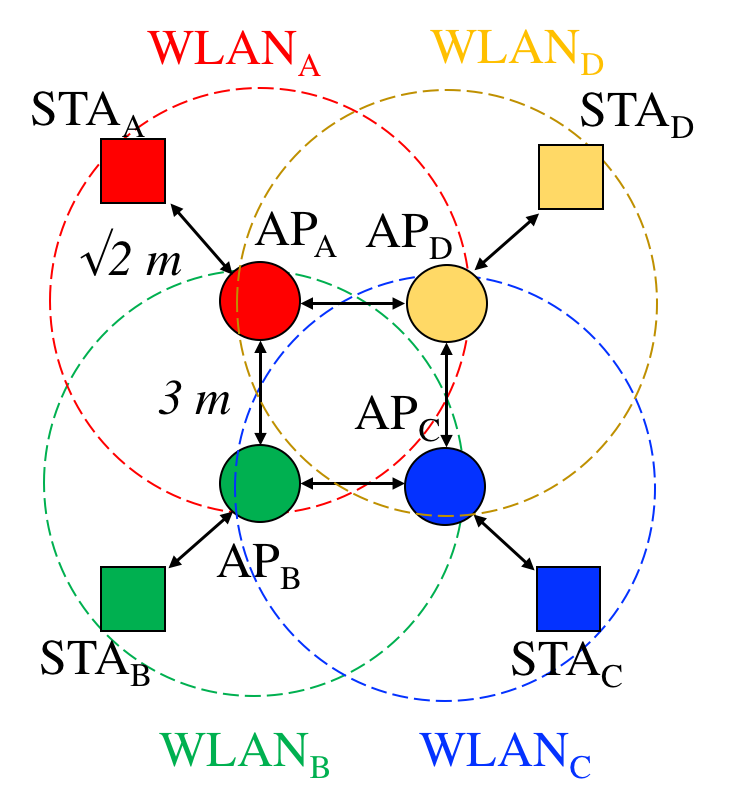}
		\caption{Scenario}
		\label{fig:selfish_s1}
	\end{subfigure}
	\begin{subfigure}[b]{0.36\textwidth}
		\includegraphics[width=\textwidth]{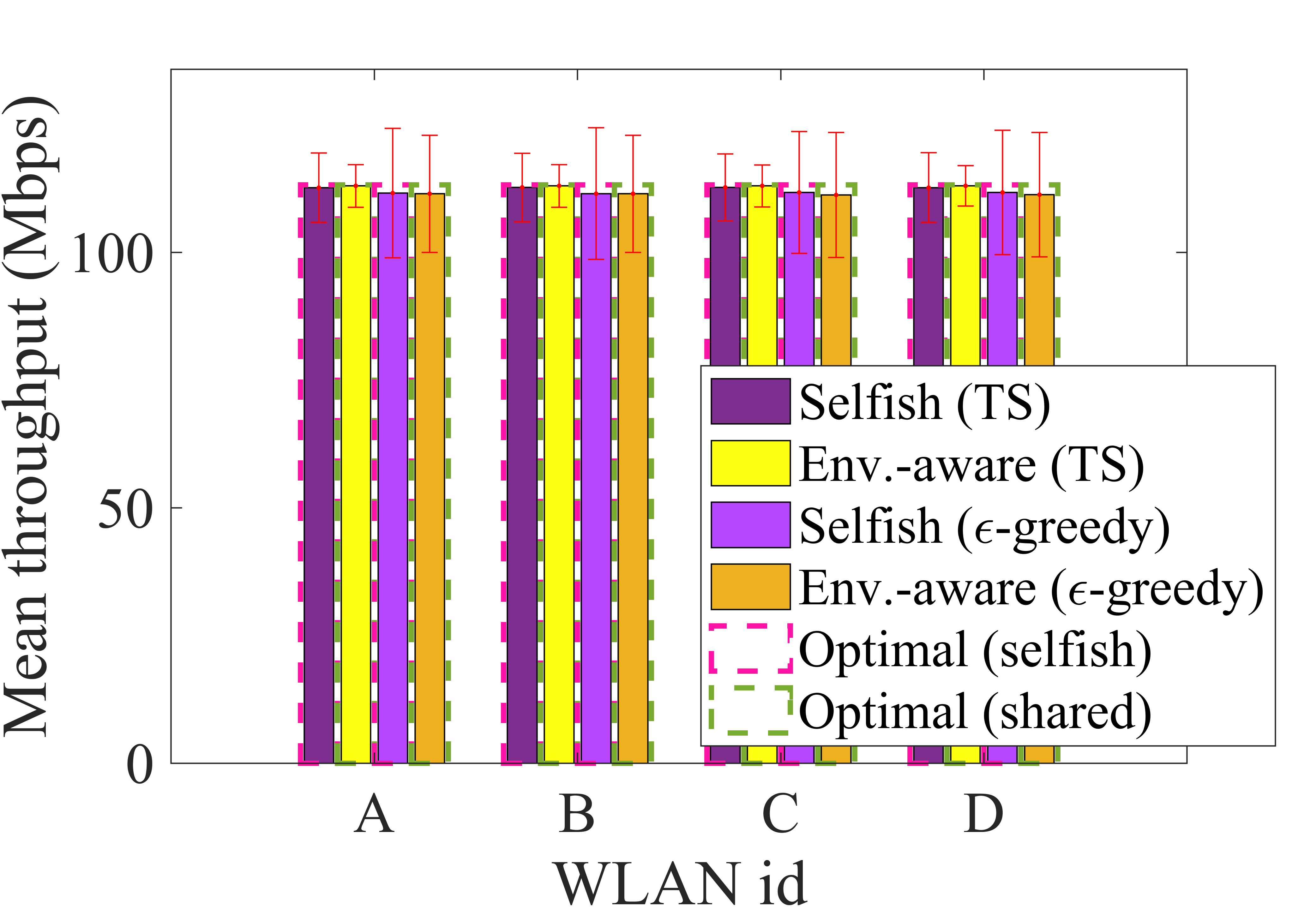}
		\caption{Mean individual throughput}
		\label{fig:4_grid_selfish_benefits_mean_tpt}
	\end{subfigure}
	\begin{subfigure}[b]{0.36\textwidth}
		\includegraphics[width=\textwidth]{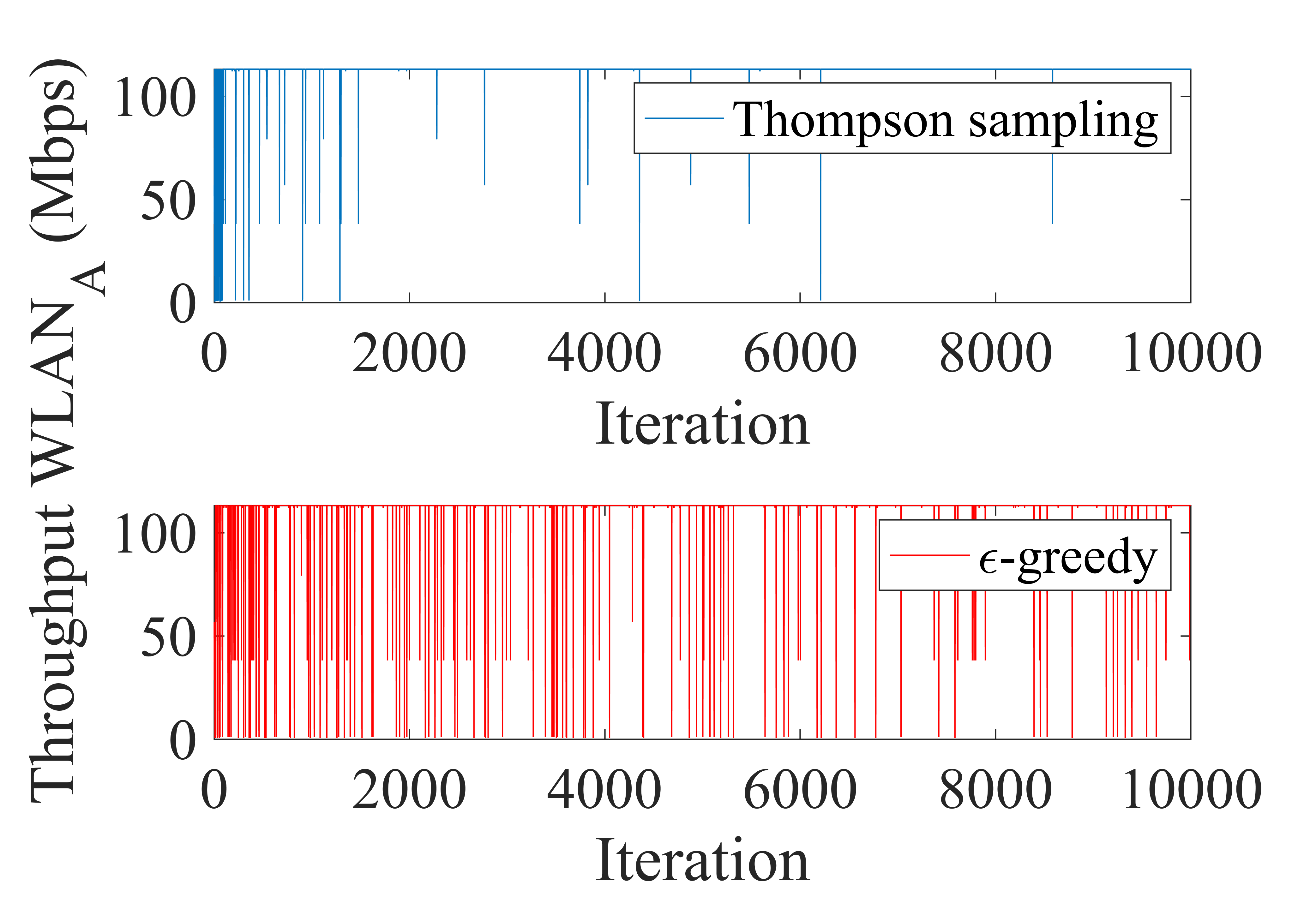}
		\caption{Throughput $\text{WLAN}_\text{A}$}
		\label{fig:experiment_2_2_variability}
	\end{subfigure}
	\caption{Potential of both selfish and environment-aware Thompson sampling (10,000 iterations are considered). (a) Scenario in which STAs are placed conservatively regarding inter-WLAN interference, (b) Average throughput per WLAN for Selfish and Environment-aware learning, both for Thompson sampling (TS) and $\varepsilon$-greedy. The standard deviation of the average throughput between iterations is shown in red. The pink and green dashed lines indicate the maximum performance achieved per WLAN regarding both selfish and environment-aware optimal solutions, respectively. (c) Temporal throughput achieved by $\text{WLAN}_\text{A}$ when learning selfishly through Thompson sampling and $\varepsilon$-greedy, respectively.}
	\label{fig:4_grid_selfish_benefits}
\end{figure}

For that, we consider a symmetric grid formed by 4 WLANs (Figure \ref{fig:selfish_s1}), which can choose from the same range of CST and transmit power levels. In this scenario, there exists an optimal configuration that can be reached by each WLAN, regardless of the others' actions. Therefore, in case all the WLANs discover the optimal action, a Nash Equilibrium is conformed, so that no individual can obtain further benefits by deviating from its strategy. As previously done, we focus on the average throughput achieved by each WLAN (Figure \ref{fig:4_grid_selfish_benefits_mean_tpt}), and the temporal throughput in $\text{WLAN}_\text{A}$ (Figure \ref{fig:experiment_2_2_variability}).

Our results show that all the WLANs are able to rapidly find the configuration that grants the maximum possible throughput, for each of the proposed learning methods. A collaborative behavior between WLANs occur despite learning selfishly. The primary reason of such a collaboration lies in the symmetries found in the scenario, and on the ability of each WLAN to compete for resources in a fair manner. Finally, and similarly to what is shown in Subsection \ref{subsubsection:fairness}, $\varepsilon$-greedy is shown to lead to a significantly higher variability than Thompson sampling.

\subsubsection{Competition Effects}
\label{subsubsection:competition}  

The scenario shown in Section \ref{subsubsection:resources_maximization} frames a conservative environment in which the inter-WLAN interference is low, i.e., APs belonging to different WLANs are distant enough, and STAs are reasonably close to their AP. However, if we refer to a less idyllic situation, applying RL may not be as effective as before. In particular, we are interested in showing the effects of using both selfish and environment-aware strategies in highly competitive environments. In those cases, it may happen that the optimal global solution becomes obfuscated because the action-selection procedure is held individually. 

\begin{figure}[h!!!!]
	\centering
	\begin{subfigure}[b]{0.3\textwidth}
		\includegraphics[width=\textwidth]{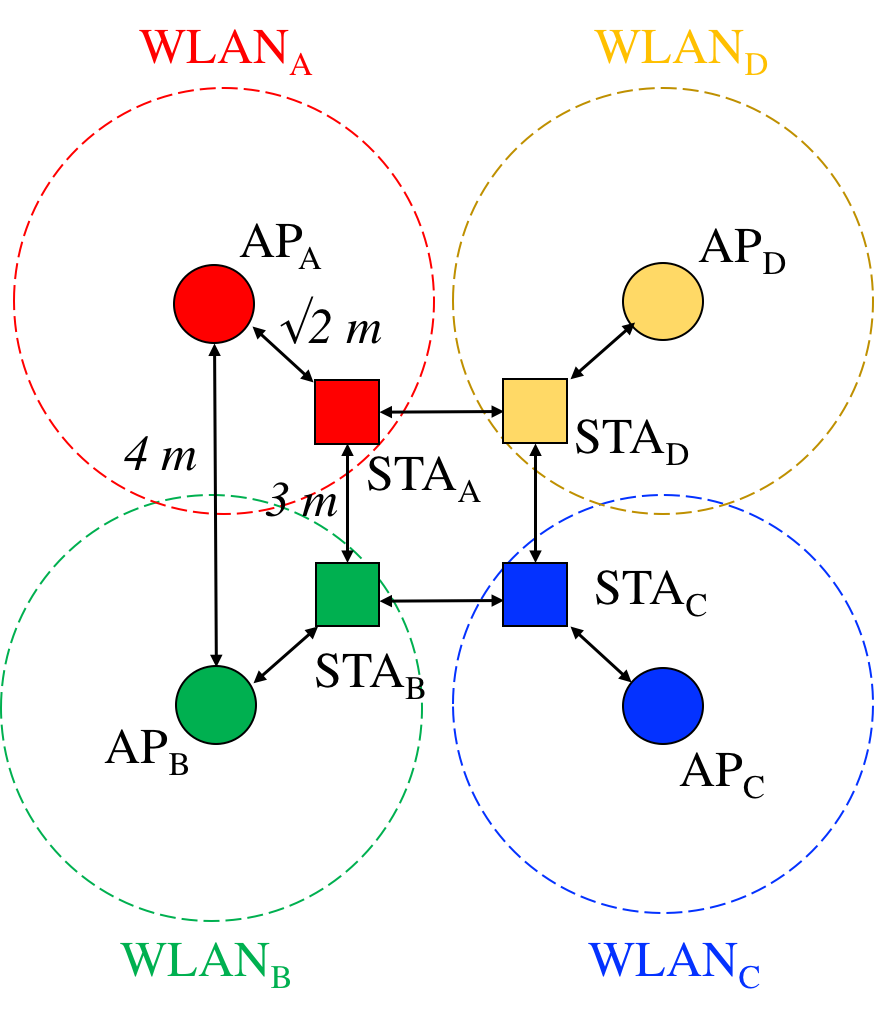}
		\caption{Scenario}
		\label{fig:adversarial_issues_scenario}
	\end{subfigure}\\
	\begin{subfigure}[b]{0.48\textwidth}
		\includegraphics[width=\textwidth]{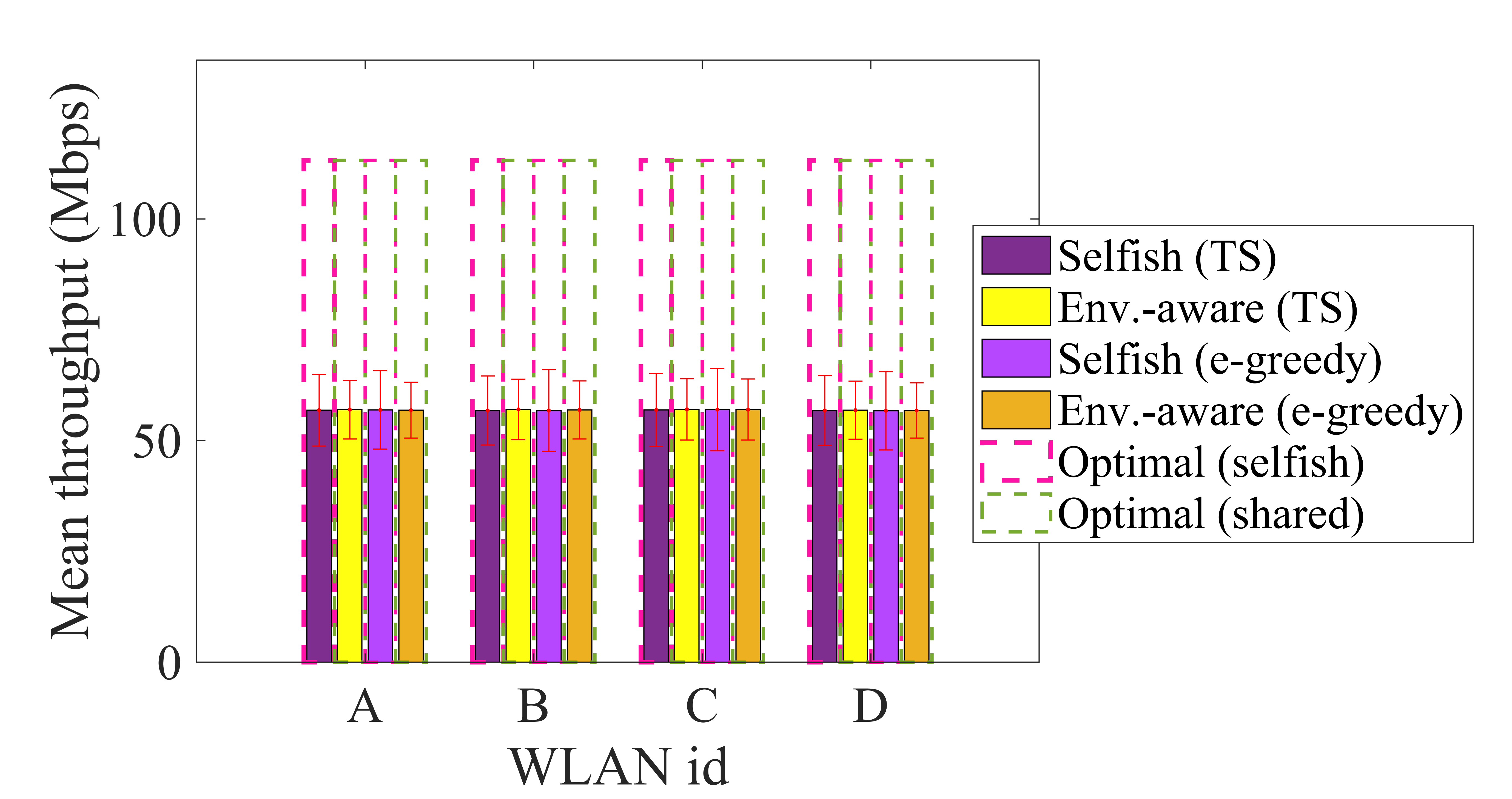}
		\caption{Mean individual throughput}
		\label{fig:4_grid_selfish_adversarial_mean_tpt}
	\end{subfigure}
	\begin{subfigure}[b]{0.36\textwidth}
		\includegraphics[width=\textwidth]{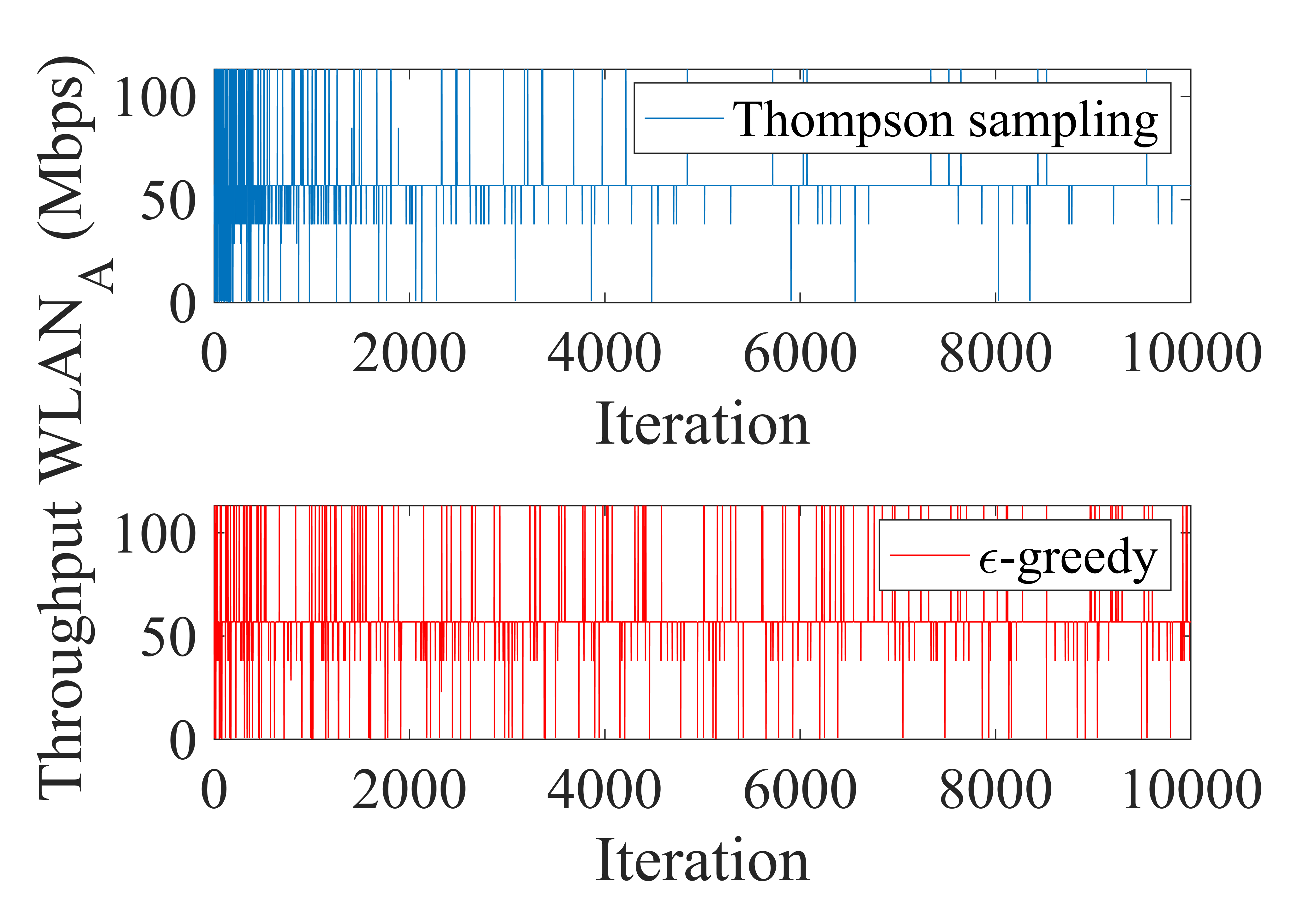}
		\caption{Throughput $\text{WLAN}_\text{A}$}
		\label{fig:experiment_2_3_variability}
	\end{subfigure}
	\caption{Competition issues in both selfish and environment-aware Thompson sampling (10,000 iterations are considered). (a) Scenario in which STAs are placed in a greedy way regarding inter-WLAN interference, (b) Average throughput per WLAN for Selfish and Environment-aware learning, both for Thompson sampling (TS) and $\varepsilon$-greedy. The standard deviation of the average throughput between iterations is shown in red. The pink and green dashed lines indicate the maximum performance achieved per WLAN regarding both selfish and environment-aware optimal solutions, respectively. (c) Temporal throughput achieved by $\text{WLAN}_\text{A}$ when learning selfishly through Thompson sampling and $\varepsilon$-greedy, respectively.}
	\label{fig:4_grid_selfish_adversarial}
\end{figure}

To show the implications of intensive competition between WLANs, let us propose the nodes distribution shown in Figure \ref{fig:adversarial_issues_scenario}. In this scenario, interactions between WLANs are more prone to generate performance issues, since all the STAs are more exposed to inter-WLAN interference. In particular, the optimal solution for both individual performance and max-min throughput is obtained only if all the WLANs use the minimum transmit power and the maximum sensitivity. The results of applying both selfish and environment-aware strategies are shown in Figures \ref{fig:4_grid_selfish_adversarial_mean_tpt} and \ref{fig:experiment_2_3_variability}. 

As it can be observed from Figure \ref{fig:4_grid_selfish_adversarial_mean_tpt}, none of the WLANs is able to reach the optimum performance, neither for the selfish nor the environment-aware reward. We identify the fact that actions are selected individually as the main cause of such a performance inefficiency. 

Regarding selfish learning, WLANs choosing the optimal configuration are more susceptible to be affected by inter-WLAN interference (asymmetries between WLANs are generated). First of all, since the optimal configuration entails using the minimum transmit power, the generated interference is minimized. As a result, the rest of WLANs can properly operate on the channel (they sense it free). However, these same WLANs cannot distinguish between the right and the harmful action from the global perspective, since both options lead to the optimal individual throughput (at the expense of harming the WLAN that is behaving properly). Henceforth, selfish WLANs are prone to act aggressively (i.e., use a high transmit power and limit the sensitivity area) in high-interference situations, which leads to obfuscate the optimal solution (even in terms of selfishness). In short, acting selfishly in this kind of scenario is not as effective as providing a certain level of collaboration that allows to identify the optimal global configuration.

Moreover, and similarly to the selfish approach, using an environment-aware metric is not enough to properly maximize the spectral efficiency. Despite WLANs act according to a joint reward, the same limitations occur due to the weakness of optimal actions in front of the environment. When a given WLAN selects the optimal action, probably it would not obtain the highest possible reward, since it is subject to the others' configuration. Since WLANs learn independently (even if the others' throughput is considered), the probabilities for all to choose the optimal action are very low. In consequence, the learning capacity is limited as for the selfish approach. To conclude, Figure \ref{fig:experiment_2_3_variability} shows that Thompson sampling is more stable than $\varepsilon$-greedy, in terms of temporal throughput variability.

\subsection{Random Scenarios}
\label{section:scalability}	
In order to further analyze the effects of applying both selfish and environment-aware strategies, we propose using 50 random scenarios, containing $\text{N} = \{2, 4, 6, 8\}$ WLANs in a $10\times10\times5$ m area (i.e., an AP every 250, 125, 83.33 and 62.5 $\text{m}^3$, respectively). WLANs are uniformly randomly distributed in the scenario, as well as STAs are randomly located between 1 and 3 meters away from their AP. Configurations are assigned so that WLANs use the same channel by default, and maximum sensitivity and transmit power. Such a configuration has been previously shown to be common in real deployments (\citealp{akella2007self}). Further details regarding the generation of random scenarios can be found in \ref{section:simulated_wireless_environment}.

Unlike in previous results, we now consider applying MABs for only 500 iterations. The main reason lies in showing the gains that can be achieved by applying MABs for short periods, i.e., before the environment significantly changes. Note that requiring large periods of time for reaching an equilibrium may not be feasible in real wireless deployments, because of the channel and users variability.

\begin{figure}[h!]
	\centering
	\begin{subfigure}[b]{0.3\textwidth}
		\includegraphics[width=\textwidth]{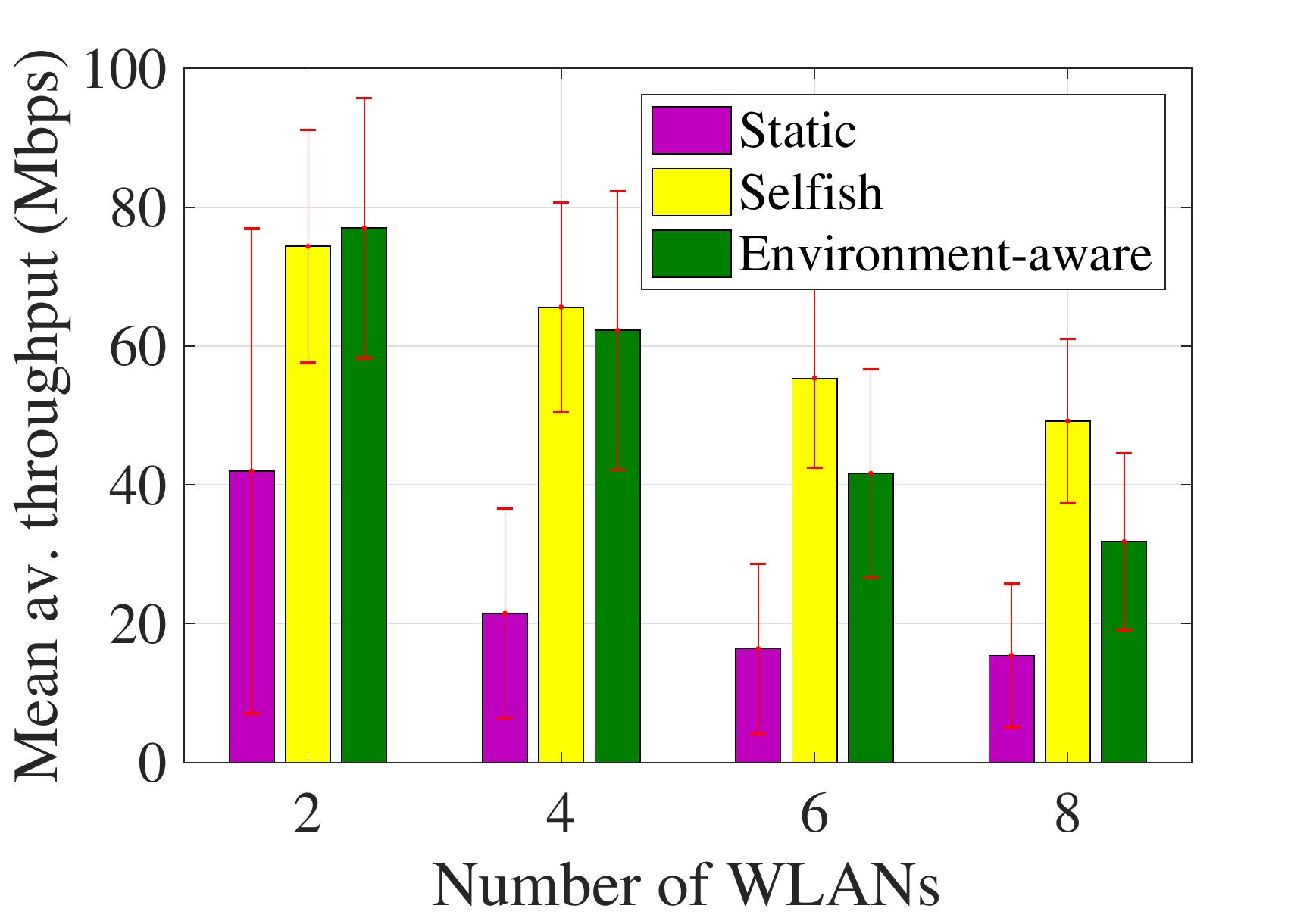}
		\caption{Average throughput}
		\label{fig:scalability_mean_tpt}
	\end{subfigure}
	\begin{subfigure}[b]{0.3\textwidth}
		\includegraphics[width=\textwidth]{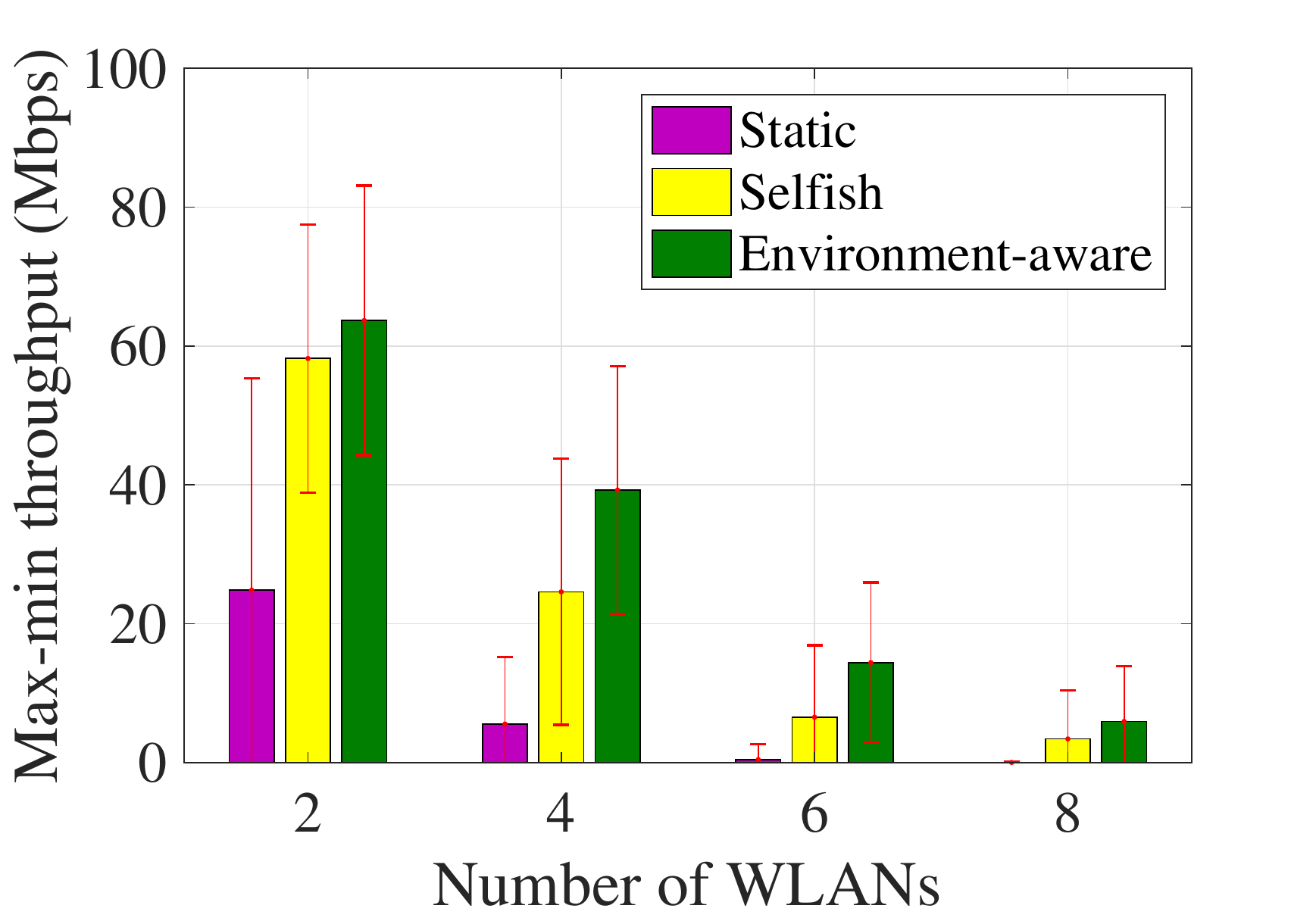}
		\caption{Max-min throughput}
		\label{fig:scalability_mean_maxmin}
	\end{subfigure}
	\begin{subfigure}[b]{0.3\textwidth}
		\includegraphics[width=\textwidth]{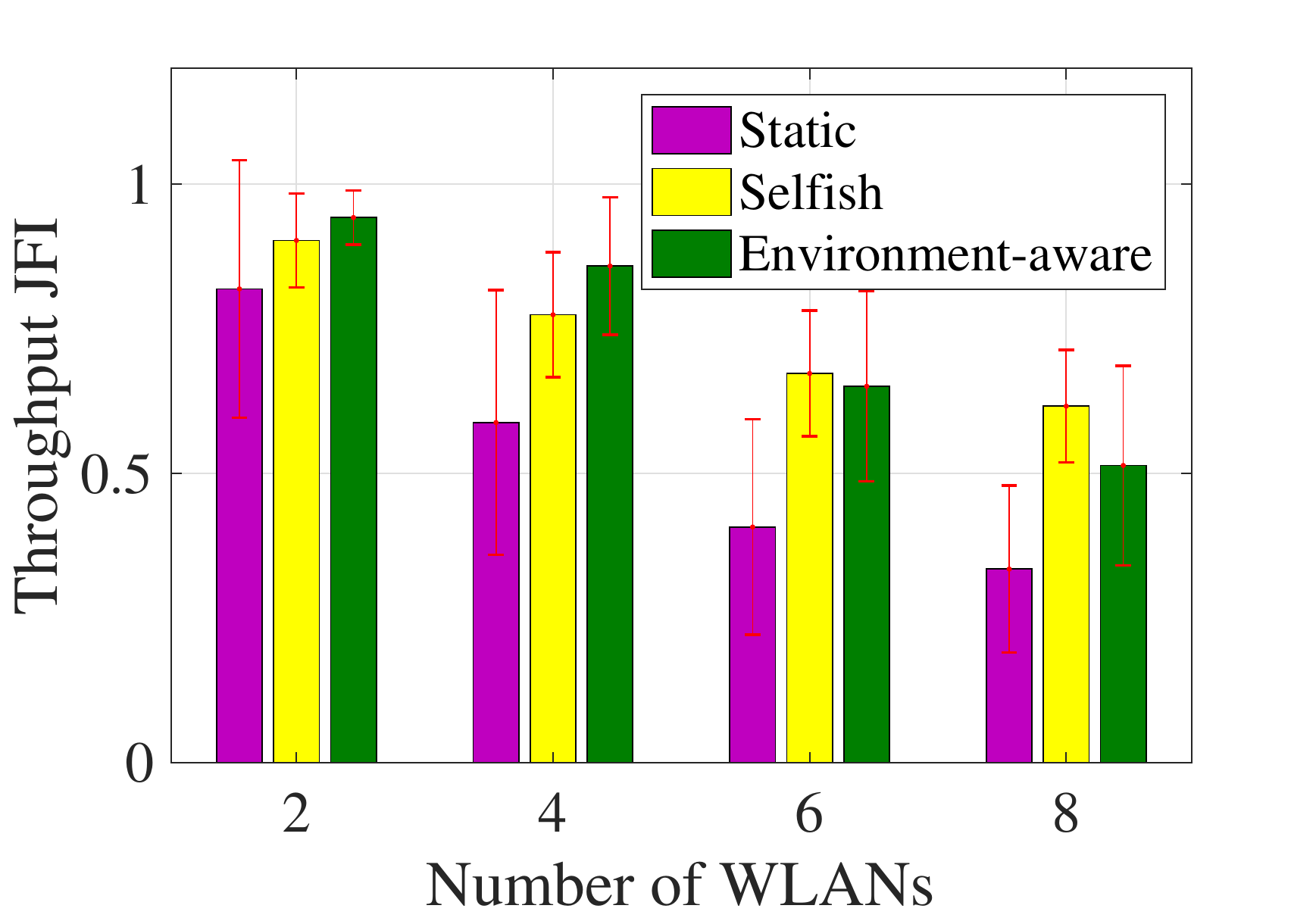}
		\caption{Throughput JFI}
		\label{fig:scalability_mean_jfi}
	\end{subfigure}
	\caption{Average results of applying 500-iteration Thompson sampling in 50 random scenarios for \{2, 4, 6, 8\} potentially overlapping WLANs. (a) Mean average throughput with standard deviation (in red) per WLAN in each scenario when using static (purple), selfish learning (yellow) or environment-aware learning (green), (b) Mean average max-min throughput with standard deviation (in red) per WLAN in each scenario when using static (purple), selfish learning (yellow) or environment-aware learning (green), (c) Mean JFI with standard deviation (in red) in each scenario when using static (purple), selfish learning (yellow) or environment-aware learning (green)}
	\label{fig:scalability_results}
\end{figure}

We first show the average results obtained by each approach in Figure \ref{fig:scalability_results}, which are compared to the static situation. The latter considers that WLANs use the initial assigned configuration. In addition, all the WLANs use the same channel (namely, channel 1). For performance evaluation, we focus on the average throughput, the max-min throughput and the JFI. Moreover, due to the impossibility of using the actual upper bound reward for each configuration, we use \emph{i)} the throughput in isolation as a maximum performance reference for the selfish strategy, and \emph{ii)} for the environment-aware strategy, the minimum throughput noticed among the individual performances of the potentially overlapping WLANs, so that their throughput in isolation is considered.

As shown, the average throughput obtained per scenario through selfish Thompson sampling outperforms the static configuration (refer to Figure \ref{fig:scalability_mean_tpt}), which is evidence of the poor spectral efficiency achieved in current deployments. In all the cases, using MABs allows to maximize the static performance. In addition, we can observe that selfish learning grants higher throughput than the environment-aware as density increases. Regarding fairness, the selfish approach is shown to work better in average (refer to Figure \ref{fig:scalability_mean_jfi}), because WLANs in a bad situation are able to self-adjust themselves in a competitive environment. However, this is not directly related to the max-min throughput, which is the goal of the environment-aware approach (refer to Figure \ref{fig:scalability_mean_maxmin}). Unfortunately, guaranteeing a certain minimum throughput to the less privileged WLANs in terms of interference becomes more challenging as the number of overlapping nodes increases. Such an issue is highly conditioned by the distance between the AP and the STA of a given WLAN.

\begin{figure}[h!]
	\centering
	\begin{subfigure}[b]{0.4\textwidth}
		\includegraphics[width=\textwidth]{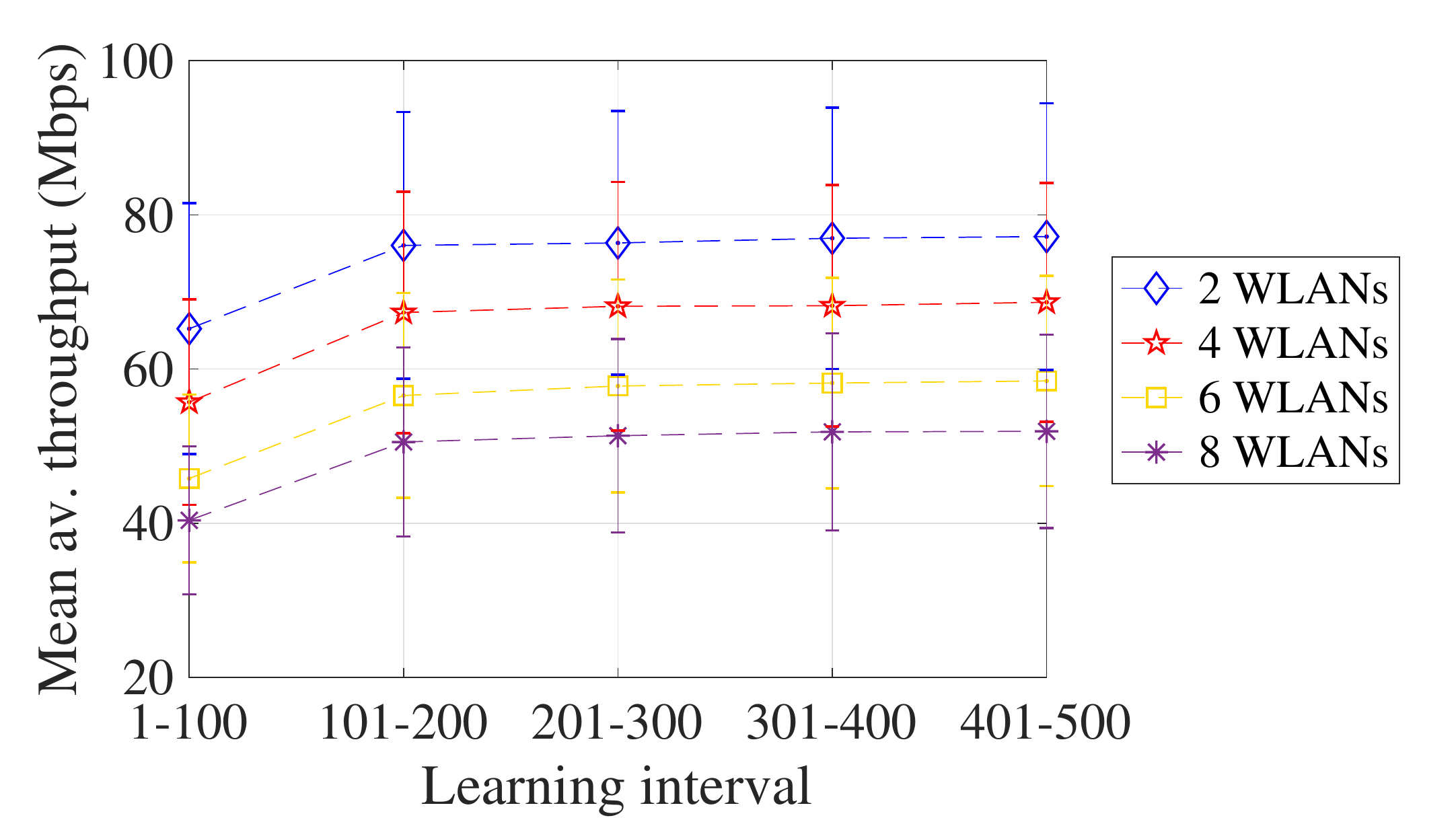}
		\caption{Selfish}
		\label{fig:scalability_selfish_accumulated}
	\end{subfigure}
	\begin{subfigure}[b]{0.4\textwidth}
		\includegraphics[width=\textwidth]{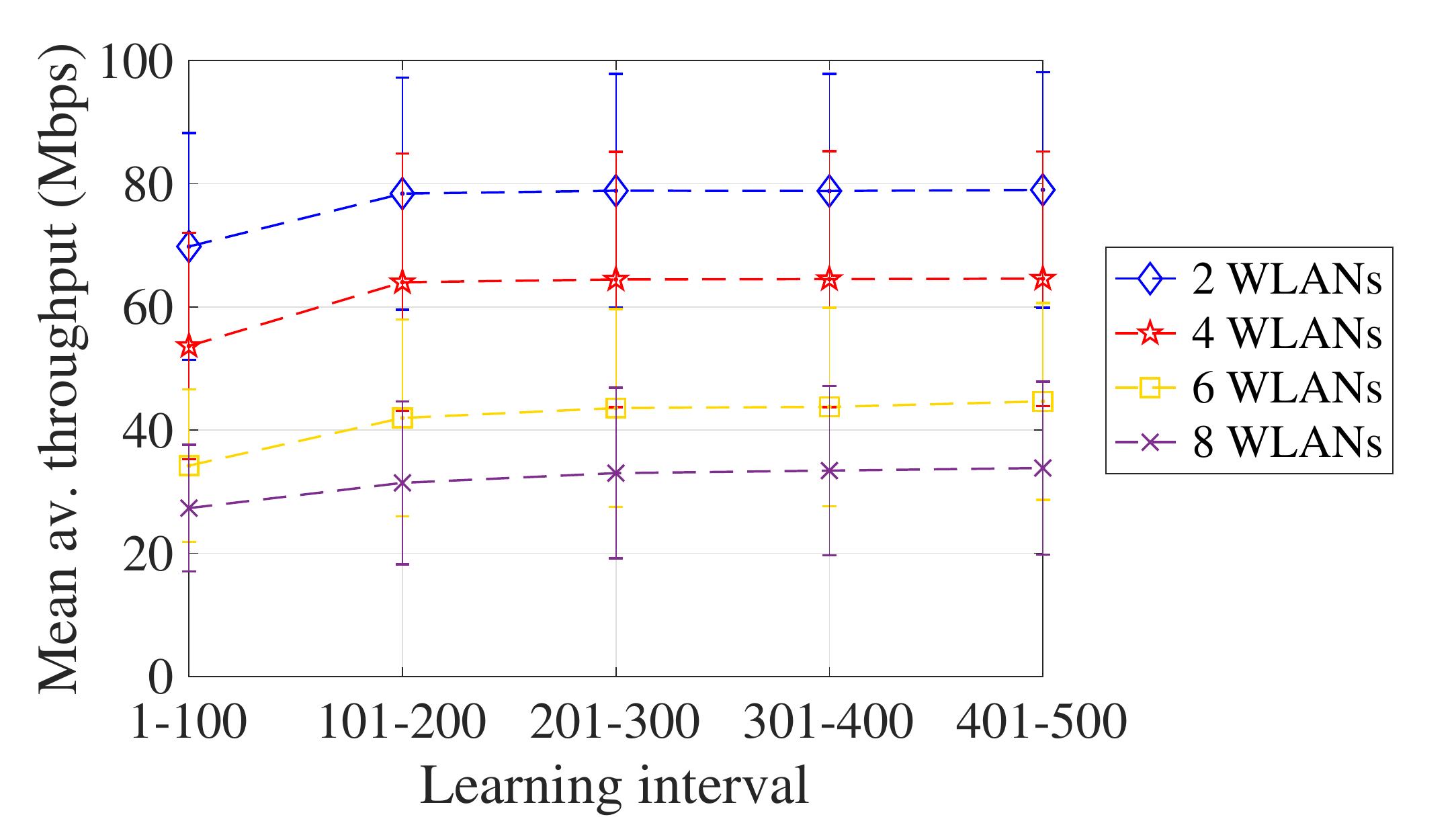}
		\caption{Environment-aware}
		\label{fig:scalability_informed_accumulated}
	\end{subfigure}
	\caption{Mean average throughput with standard deviation achieved during specific intervals. 500 iterations are considered for both selfish and environment-aware Thompson sampling in 50 random scenarios for \{2, 4, 6, 8\} overlapping WLANs. (a) Results for the selfish strategy, (b) Results for the environment-aware strategy.}
	\label{fig:cumulative_tpt}
\end{figure}

Finally, to further illustrate the enhancements achieved by applying learning, we plot the average throughput obtained for different learning phases in Figure \ref{fig:cumulative_tpt}. By showing the performance experienced for each interval of 100 iterations, we aim to emphasize on the progressive gains achieved along the learning operation. As previously mentioned, wireless environments are highly varying, thus a fast convergence is essential for any learning algorithm.

For both strategies, we observe a big gain experienced during the first two intervals, which becomes stable from that point onwards. Thus, even if an equilibrium is not reached, a significant increase of the average performance is rapidly experienced. Regarding environment-aware learning, a greater enhancement is provided when density is low (two and four WLANs). However, when density increases, selfish learning achieves a higher average performance earlier. The key reason lies in the fact that max-min throughput is more likely to be low as the number of overlapping devices increases. Therefore, the learning procedure is slowed down due to the impossibility of finding an appropriate solution that alleviates the poor performance achieved by the most vulnerable WLANs. In opposite, learning selfishly speeds up performance trading fairness off.

\section{Conclusions}
\label{section:conclusions}		
In this work, we addressed the potential and feasibility of applying decentralized online learning to wireless networks, as a contribution to the debate about whether future WLANs should remain decentralized or evolve towards centralized mechanisms (such as in cellular networks). To that purpose, we delved into the SR problem in IEEE 802.11 WLANs and presented a practical learning-based application to overcome it. In particular, we modeled the problem through MABs and showed two strategies based on the Thompson sampling action-selection method. The first one is based on learning selfishly, where the reward that an agent obtains after playing an action is granted according to its own throughput, regardless of the performance achieved by the overlapping WLANs. The second one, referred to as environment-aware learning, quantifies how good actions are based on the max-min throughput achieved in the network. This can be done by inferring the performance obtained by the surrounding WLANs. 

By using the SR problem as a guiding thread, we analyzed the main considerations that must be done when applying decentralized learning methods to wireless communications problems. Among them, we highlighted practical issues such as convergence assessment or the difficulties on developing an appropriate reward generation system. Finally, we evaluated two learning policies in terms of fairness and throughput, so as to show their potential and major implications. Despite learning selfishly has been shown to generate unfair situations, its potential at maximizing the aggregate performance is very promising in certain scenarios. Moreover, even if environment-aware methods allow to solve fairness-related issues, the fact of learning in a decentralized way is not a guarantee for finding the best-performing configuration. In addition, environment-aware learning may severely limit the aggregate performance in benefit of few WLANs. 

As a final conclusion, we remark the potential of applying uncoordinated MABs in dense WLANs, thus bringing hope for decentralized deployments in front of centralized systems. However, for practical application, such a kind of mechanisms are required to take the environment into consideration, since selfish approaches are prone to generate unfair situations. Therefore, other important challenges such as inter-WLANs communication must be overcome. Moreover, the utilization of collaborative approaches raises several questions regarding fairness ascertainment. For instance, is it worth to significantly reduce the performance of many WLANs in benefit to less privileged ones in terms of location?

Future work will also consider the use of beamforming to improve spatial reuse. By defining multiple beamforming sectors (i.e., 4 or 8), multiple simultaneous transmissions can be performed from different APs if they select non-interfering sectors for transmitting. Thus, it adds another degree of freedom which combined with transmission power, CCA, and channel allocation adaptation, which may be done per sector, could further contribute to improve the overall system performance.

\section*{Acknowledgment}
This work has been partially supported by the Spanish Ministry of Economy and Competitiveness under the Maria de Maeztu Units of Excellence Programme (MDM-2015-0502), by the Catalan Government SGR grant for research support (2017-SGR-1188), by the European Regional Development Fund under grant TEC2015-71303-R (MINECO/FEDER), and by a Gift from the Cisco University Research Program (CG\#890107, Towards Deterministic Channel Access in High-Density WLANs) Fund, a corporate advised fund of Silicon Valley Community Foundation.

The authors would like to thank the anonymous reviewers, which thorough work and insightful comments were of great help to improve this paper.

\begin{appendices}
\section{Wireless Environment}
\label{section:simulated_wireless_environment}	
Here we provide details on the wireless environment used to simulate IEEE 802.11 WLANs behavior. First, physical medium effects are modeled by following the specification provided in the IEEE 802.11ax standard for residential scenarios (\citealp{merlin2015tgax}), which includes specific path-loss and shadowing models. We have chosen this scenario because it is very representative for next-generation dense and chaotic deployments. In particular, since we refer to random scenarios, we capture the essence of the 11ax residential scenario by proposing a model that takes into account the walls and floor frequencies, rather than the actual location of walls and floors. Accordingly, the power loss $\text{PL}_d$ in such an environment is given by:
\begin{align}
	\text{PL}_d = 40.05 + 20 \log_{10}\Big(\frac{f_c}{2.4}\Big) + 20 \log_{10}(\min(d,5)) + \nonumber I_{d>5} \cdot 35\log_{10}\Big(\frac{d}{5}\Big) + 18.3 F^{\frac{F+2}{F+1}-0.46} + 5W
	\nonumber
\end{align}	
where $f_c$ is the frequency in GHz, $d$ is the distance between the transmitter and the receiver in meters, and $F$ and $W$ are the average number of floors and walls traversed per meter, respectively. The original scenario considers a 5-floor building, with twenty apartments of $10\times10\times3$ meters size per floor. Regarding adjacent channel interference, we consider that consecutive channels are non-overlapping. 

Note, as well, that the data rate at which a transmitter sends data is subject to the signal strength in the receiver, which in this work is assumed to be known. IEEE 802.11ax parameters are used, so that modulations range from BPSK to 1024-QAM (\citealp{ieee2015qam}). Table \ref{tbl:simulation_parameters} details the parameters used, which include PHY and MAC specifications (\citealp{tgax2017draft}).
\begin{table*}[h!]
	\centering
	\resizebox{\textwidth}{!}{\begin{tabular}{|l|l|l|}
		\hline
		\textbf{Parameter} & \textbf{Description} & \textbf{Value} \\ \hline
		$\mathcal{C}$ & Set of channels & 1 / 2 \\ \hline
		$\mathcal{S}$ & Set of sensitivity thresholds & -68 dBm / -90 dBm \\ \hline
		$\mathcal{T}$ & Set of transmit power values & 5 dBm / 20 dBm \\ \hline
		$d_{\text{AP,STA}}^{min}$ / $d_{\text{AP,STA}}^{max}$& Min/max distance AP - STA & 1 m  / 3 m\\ \hline
		$(x, y, z)$ & 3D map dimensions in each axis & (10, 10, 5) m\\ \hline
		W & Channel bandwidth & 20 MHz \\ \hline
		$F$ & Central frequency & 5 GHz \\ \hline
		$\text{SUSS}$ & Spatial streams per user & 1 \\ \hline
		$G_{tx}$ & Transmitting gain & 0 dBi \\ \hline
		$G_{rx}$ & Reception gain & 0 dBi \\ \hline
		N & Floor noise level & -95 dBm \\ \hline
		CE & Capture Effect threshold & 10 dBm \\ \hline
		$T_s$ & Symbol duration & 9 $\mu$s \\ \hline
		DIFS/SIFS & DIFS and SIFS duration & 34 $\mu$s / 16 $\mu$s \\ \hline
		$\text{CW}_{min} / \text{CW}_{max}$ & Min/max contention window & 16 / 16 \\ \hline	
		$N_{agg}$ & Number of packets aggregated & 64 \\ \hline
		$L_{\text{DATA}}$ & Length of a data packet & 12000 bits \\ \hline
		$L_{\text{RTS}}$ / $L_{\text{CTS}}$ & Length RTS and CTS packets & 160 bits / 112 bits \\ \hline
		$L_{\text{MAC}}$ & Length MAC header & 272 bits \\ \hline
		$L_{\text{SF}}$ & Length Service Field (SF) & 16 bits \\ \hline
		$L_{\text{MPDU}}$ & MPDU delimiter & 32 bits \\ \hline
		$L_{\text{Tail}}$ & Length tail & 6 bits \\ \hline
		$L_{\text{BACK}}$ & Length block ACK & 240 bits \\ \hline	
		$T_{\text{RTS}}$ & RTS packet duration & $20\cdot 10^{-6} + \frac{L_{\text{SF}}+L_{\text{RTS}}+L_{\text{Tail}}}{R}T_s$ s\\ \hline	
		$T_{\text{CTS}}$ & CTS packet duration & $20\cdot 10^{-6} + \frac{L_{\text{SF}}+L_{\text{CTS}}+L_{\text{Tail}}}{R}T_s$ s \\ \hline	
		$T_{\text{DATA}}$ & Data packet duration & $36\cdot 10^{-6} + \text{SUSS} \cdot 16 \cdot 10^{-6} + \frac{(L_{\text{SF}}+N_{agg}\cdot(304+L_{\text{DATA}})+L_{Tail})}{R}T_s$ s \\ \hline	
		$T_{\text{BACK}}$ & Block ACK duration & $20 \cdot 10^{-6} + \frac{L_{\text{SF}}+L_{\text{BACK}}+L_{\text{Tail}}}{R}T_s $ s\\ \hline	
		$\mathcal{T}$ & Traffic model & Full buffer (downlink) \\ \hline	
	\end{tabular}}
	\caption{Simulation parameters}
	\label{tbl:simulation_parameters}
\end{table*}	
\end{appendices}

\newpage
\section*{\refname}
\bibliography{references}

\end{document}